%% file: main.tex
\documentclass{article} 
\usepackage{iclr2025_conference,times,url}
\usepackage{amsmath}
\usepackage{amssymb}
\usepackage{amsthm}
\usepackage{amsfonts}
\input{prestuff}

\usepackage{graphicx}
\usepackage{hyperref,multirow,xspace,algorithm2e}
\usepackage{url}
\usepackage{natbib}

\newtheorem{theorem}{Proposition}
\newtheorem{definition}{Definition}

\usepackage{tabu}

\theoremstyle{remark}

\newcommand{\system}{{\sc Rapid}\xspace}
\newcommand{\dpdm}{{\sc DpDm}\xspace}
\newcommand{\dpldm}{{\sc Dp-Ldm}\xspace}
\newcommand{\dpsda}{{\sc Dpsda}\xspace}

\newcommand{\rev}[1]{\textcolor{blue}{#1}\xspace}

\title{RAPID: Retrieval Augmented Training of Differentially Private Diffusion Models}

\author{
Tanqiu Jiang$^\dagger$\quad
Changjiang Li$^\dagger$\quad
Fenglong Ma$^\star$\quad
Ting Wang$^\dagger$\\
$^\dagger$Stony Brook University\quad
$^\star$Pennsylvania State University
}

\iclrfinalcopy 
\begin{document}

\maketitle

\input{abstract}

\input{intro}
\input{literature}

\input{method}

\input{evaluation}

\input{conclusion}

\section*{Acknowledgment}
This work is partially supported by the National Science Foundation under Grant No. 2405136, 2406572, and 2212323.


\bibliographystyle{iclr2025_conference}
\bibliography{references}  

\newpage
\input{appendix}

\end{document}

%% file: prestuff.tex
\usepackage{epsfig,amsmath,amsfonts,epsfig,multirow,makecell,caption,soul,csquotes,color,wrapfig,subcaption,mathtools,bm,spverbatim,booktabs,tcolorbox,diagbox}
\usepackage[e]{esvect}

\usepackage{caption}
\makeatletter
\newif\if@restonecol
\makeatother

\usepackage[boxed, ruled, vlined, linesnumbered]{algorithm2e}
\SetKwRepeat{Do}{do}{while}

\newenvironment{changemargin}[2]{\begin{list}{}{
	\setlength{\topsep}{0pt}\setlength{\leftmargin}{0pt}
	\setlength{\rightmargin}{0pt}
	\setlength{\listparindent}{\parindent}
	\setlength{\itemindent}{\parindent}
	\setlength{\parsep}{0pt plus 1pt}
	\addtolength{\leftmargin}{#1}\addtolength{\rightmargin}{#2}
	}\item}
	{\end{list}}

\newenvironment{mitemize}{
	\begin{changemargin}{-10pt}{-0cm}
	\vspace{-10pt}
	\hspace{-8pt}
	\begin{itemize}
	\setlength{\itemsep}{3pt}}
	{\end{itemize}
	\vspace{2pt}
	\end{changemargin}}

\usepackage[first=0,last=9]{lcg}
\usepackage{colortbl}
\definecolor{Gray}{gray}{0.8}
\colorlet{Red}{red!10!white}
\colorlet{Blue}{blue!10!white}

\usepackage{hyperref}

\newcommand{\msec}[1]{\S\ref{#1}}
\newcommand{\mref}[1]{\,\ref{#1}}
\newcommand{\meq}[1]{Eq.\,\ref{#1}}

\newcommand{\mct}[1]{({\em #1})}

\newtcolorbox{mtbox}[1]{left=0.25mm, right=0.25mm, top=0.25mm, bottom=0.25mm, sharp corners, colframe=red!50!black, boxrule=0.5pt, title={#1}, fonttitle=\bfseries, coltitle=red!50!black, attach title to upper={\ --\ }}

\usepackage{scalerel}[2016/12/29]

\makeatletter
\providecommand{\leadsfrom}{%
  \mathrel{\mathpalette\reflect@squig\relax}%
}
\newcommand{\reflect@squig}[2]{%
  \reflectbox{$\m@th#1\leadsto$}%
}
\makeatother

\def\eqref#1{equation~\ref{#1}}

\def\1{\bm{1}}

\def\rveps{\bm{\epsilon}}

\def\rg{{\textnormal{g}}}

\def\rvg{{\mathbf{g}}}

\def\rvx{{\mathbf{x}}}

\def\rvz{{\mathbf{z}}}

\DeclareMathAlphabet{\mathsfit}{\encodingdefault}{\sfdefault}{m}{sl}
\SetMathAlphabet{\mathsfit}{bold}{\encodingdefault}{\sfdefault}{bx}{n}

\def\gB{{\mathcal{B}}}

\def\gD{{\mathcal{D}}}

\def\gK{{\mathcal{K}}}

\def\gM{{\mathcal{M}}}
\def\gN{{\mathcal{N}}}

\def\gR{{\mathcal{R}}}

\def\gU{{\mathcal{U}}}

\def\sE{{\mathbb{E}}}

\newcommand{\E}{\mathbb{E}}



%% file: abstract.tex
\begin{abstract} \label{abstract}
Differentially private diffusion models (DPDMs) harness the remarkable generative capabilities of diffusion models while enforcing differential privacy (DP) for sensitive data. However, existing DPDM training approaches often suffer from significant utility loss, large memory footprint, and expensive inference cost, impeding their practical uses.  
To overcome such limitations, we present \system\footnote{\system: \ul{R}etrieval-\ul{A}ugmented \ul{P}r\ul{I}vate \ul{D}iffusion model.}, a novel approach that integrates retrieval augmented generation (RAG) into DPDM training. Specifically, \system leverages available public data to build a knowledge base of sample trajectories; when training the diffusion model on private data, \system computes the early sampling steps as queries, retrieves similar trajectories from the knowledge base as surrogates, and focuses on training the later sampling steps in a differentially private manner. Extensive evaluation using benchmark datasets and models demonstrates that, with the same privacy guarantee, \system significantly outperforms state-of-the-art approaches by large margins in generative quality, memory footprint, and inference cost, suggesting that retrieval-augmented DP training represents a promising direction for developing future privacy-preserving generative models. The code is available at: \url{https://github.com/TanqiuJiang/RAPID}.
\end{abstract}

%% file: intro.tex
\section{Introduction} \label{Introduction}

The recent advances in diffusion models have led to unprecedented capabilities of generating high-quality, multi-modal data~\citep{ho2020denoising,kong2020diffwave,bar2024lumiere}. However, training performant diffusion models often requires massive amounts of training data, raising severe privacy concerns in domains wherein data is sensitive. For instance,  \citet{carlini2023extracting} show that compared with other generative models, diffusion models are especially vulnerable to membership inference attacks, due to their remarkable modeling capabilities; meanwhile, \citet{wen2023detecting} show that text-conditional diffusion models trained with text-image pairs can produce images almost identical to certain training samples with proper prompting.



This pressing need has spurred intensive research on enforcing privacy protection in diffusion model training. Notably, \citet{dockhorn2022differentially} proposed the concept of differentially private diffusion models (DPDMs), which incorporate DP-SGD~\citep{abadi2016deep} into diffusion model training, providing guaranteed privacy; \citet{ghalebikesabi2023differentially} further applied DP during diffusion model fine-tuning, first pre-training a denoising diffusion probabilistic model~\citep{ho2020denoising} on public data and then fine-tuning the model on private data under DP constraints. Similarly, \citet{lyu2023differentially} employed the same strategy but extended it to latent diffusion models~\citep{rombach2022high}. However,  existing methods suffer from major limitations. \mct{i} Significant utility loss -- The quality of generated samples often drops sharply under tightened privacy budgets. For instance, \dpdm~\citep{dockhorn2022differentially} fails to synthesize recognizable images on CIFAR10 under a DP budget of $\epsilon = 1$. \mct{ii} Large memory footprint -- To reduce the noise magnitude applied at each iteration, most approaches adopt excessive batch sizes (e.g., $B$ = 8,192 samples per batch). As common DP frameworks (e.g., {\sc Opacus}~\citep{opacus}) have peak memory requirements of $\mathcal{O}(B^2)$, this severely limits the size of usable diffusion models. \mct{iii} Expensive inference cost -- Similar to non-private diffusion models, existing methods require expensive, iterative sampling at inference, impeding them from synthesizing massive amounts of data.

To address such challenges, we present \system, a novel approach that integrates retrieval augmented generation (RAG)~\citep{lewis2020retrieval,blattmann2022retrieval} into DPDM training. Our approach is based on the key observation that small perturbations to the early sampling steps of diffusion models have a limited impact on the overall sampling trajectory~\citep{sensitivity}. This allows us to reuse previously generated trajectories, if similar to the current one, as effective surrogates to reduce both training and inference costs~\citep{zhang2023redi}. Leveraging this idea, \system utilizes available public data to pre-train a diffusion model and build a knowledge base of sampling trajectories. It further refines the model using private data: it first computes the early sampling steps, retrieves similar trajectories from the knowledge base as surrogates, and focuses on training the later sampling steps under DP constraints. Compared to prior work, \system offers several major advantages: 

\begin{mitemize}
    \item  It achieves a more favorable privacy-utility trade-off by fully utilizing public data and only training the later sampling steps on private data;
\item It significantly reduces batch-size requirements by leveraging the retrieved sample trajectories, making the use of large diffusion models feasible; 
\item It greatly improves inference efficiency by skipping intermediate sampling steps via RAG. 
\end{mitemize}

Extensive evaluation using benchmark datasets and models demonstrates that \system outperforms state-of-the-art methods by large margins in there key areas: \mct{i} generative quality (e.g., improving FID score to 63.2 on CIFAR10 under a DP budget $\epsilon = 1$), \mct{ii} memory footprint (e.g., reducing the required batch size to just 64 samples per batch), and \mct{iii} inference efficiency (e.g., saving up to 50\% of the inference cost). Our findings suggest that integrating RAG into DP training represents a promising direction for developing future privacy-preserving generative models.

%% file: literature.tex
\section{Related Work}

{\bf Differentially private data generation.} As an important yet challenging problem, enforcing DP into training a variety of advanced generative models has attracted intensive research effort~\citep{sok}, including generative adversarial networks~\citep{gan,gs-wgan,pate-gan}, variational autoencoders~\citep{jiang2022dp2vae}, and customized architectures~\citep{liew2022pearl,vinaroz2022hermite,harder2021dpmerf}. For instance, \citet{harder2023pretrained} pre-train perceptual features using public data and fine-tune only data-dependent terms using maximum mean discrepancy under the DP constraint. 

{\bf Differentially private diffusion models.} In contrast, the work on privatizing diffusion models is relatively limited. Notably, \dpdm~\citep{dockhorn2022differentially} integrate DP-SGD~\citep{abadi2016deep} with a score-based diffusion model~\citep{song2021scorebased}; \citet{ghalebikesabi2023differentially} propose to pre-train a diffusion model with public data and then fine-tune the model using DP-SGD on private data; \dpldm~\citep{lyu2023differentially} apply a similar fine-tuning strategy to a latent diffusion model~\citep{rombach2022high}; PrivImage~\citep{privimage} queries the private data distribution to select semantically similar public samples for pretraining, followed by DP-SGD fine-tuning on the private data. However, all these methods share the drawbacks of significant utility loss, excessive batch sizes, or expensive inference costs. Beyond the pre-training/fine-tuning paradigm, recent work also explores synthesizing DP datasets by querying commercial image generation APIs (e.g., Stable Diffusion and DALL-E2) to approximate the distributions of private data~\citep{wangdp,dpsda}.

{\bf Retrieval augmented generation.} Initially proposed to enhance the generative quality of NLP models by retrieving related information from external sources~\cite{khandelwal2019generalization,lewis2020retrieval,guu2020retrieval}, RAG has been extended to utilize local cohorts in the training data to facilitate image synthesis. For instance, rather than directly outputting the synthesized sample, \citet{casanova2021instance} compute the average of the sample's nearest neighbors in the training data. \citet{blattmann2022retrieval} use an external image dataset to provide enhanced conditional guidance, augmenting the text prompt during the text-to-image generation. \cite{zhang2023redi} explore RAG to accelerate the inference process of a diffusion model by reusing pre-computed sample trajectories as surrogates for skipping intermediate sampling steps. However, existing work primarily focuses on employing RAG in the inference stage to improve generative quality or efficiency.


To our best knowledge, this presents the first work on integrating RAG in the DP training of diffusion models, aiming to improve the generative quality, memory footprint, and inference efficiency over the existing DPDM approaches.

%% file: method.tex
\section{RAPID}

Next, we present \system, a novel approach for training differentially private diffusion models by leveraging retrieval-augmented generation.

\begin{figure}[!t]
    \centering
    \includegraphics[width=0.72\textwidth]{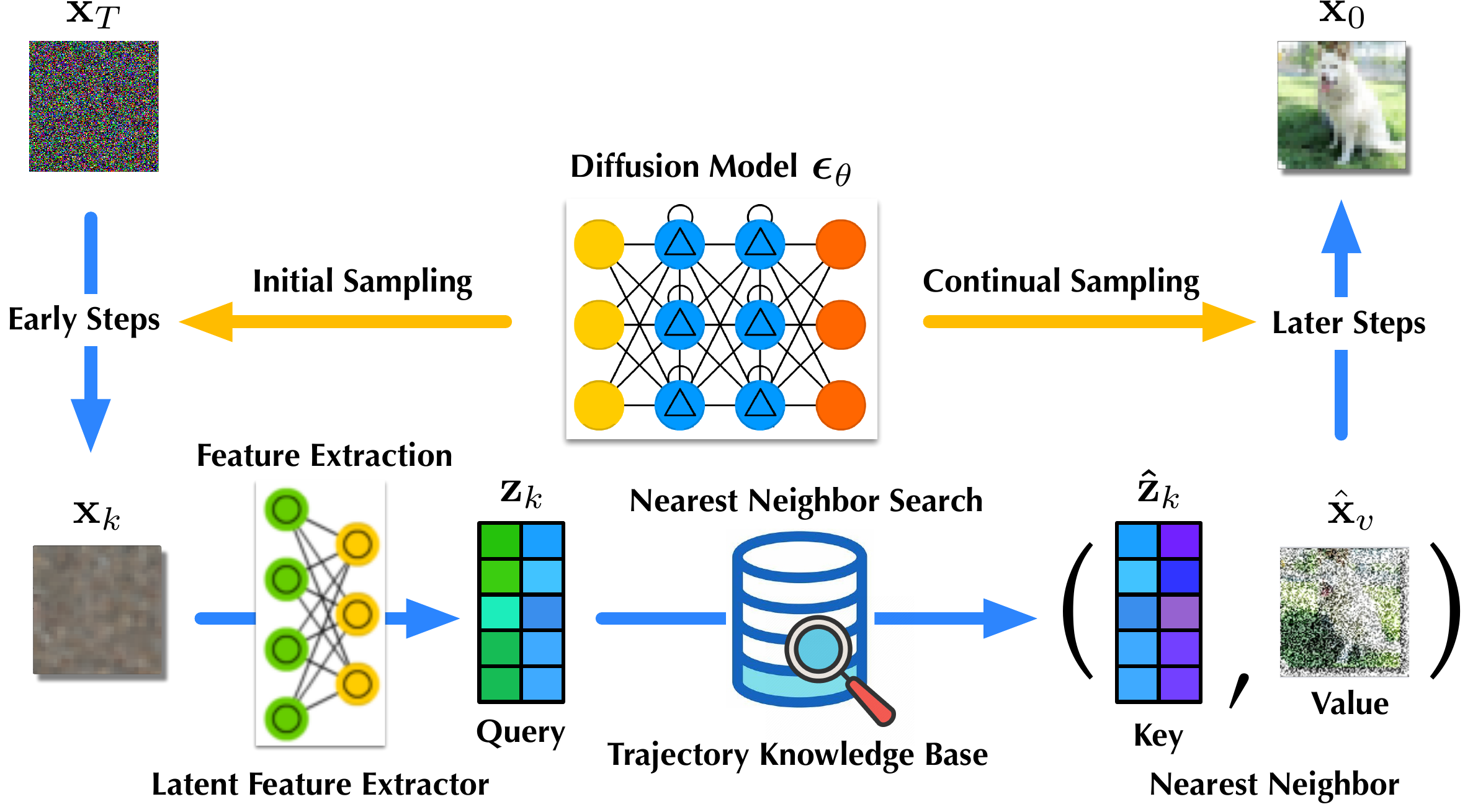}
        \caption{Overall framework of \system.}
    \label{fig:NNflow}
\end{figure}

\subsection{Preliminaries}

A diffusion model consists of a forward diffusion process that converts original data $\rvx_0$ to its latent $\rvx_t$ (where $t$ denotes the timestep) via progressive noise addition and a reverse sampling process that starts from latent $\rvx_t$ and generates data $\rvx_0$ via sequential denoising steps. 

Take the denoising diffusion probabilistic model (DDPM)~\citep{ho2020denoising} as an example. Given $\rvx_0$ sampled from the real data distribution $q_\mathrm{data}$, the diffusion process is formulated as a Markov chain: 
\begin{equation}
\label{eq:sampling}
q(\rvx_t | \rvx_{t-1}) = \gN(\rvx_t ; \sqrt{1 - \beta_t}\rvx_{t-1}, \beta_t \mathbf{I}) 
\end{equation}
where $\{\beta_t\in (0,1)\}_{t=1}^T$ specifies the variance schedule. For sufficiently large $T$, the latent $\rvx_T$ approaches an isotropic Gaussian distribution. Starting from $p(\rvx_T) = \gN(\rvx_T; \bm{0}, \bm{\text{I}})$, the sampling process maps latent $\rvx_T$ to data $\rvx_0$ in $q_\mathrm{data}$ as a Markov chain with a learned Gaussian transition:
\begin{equation}
p_\theta(\rvx_{t-1} | \rvx_t) = \gN( \rvx_{t-1}; \bm{\mu}_\theta(\rvx_t, t), \bm{\Sigma}_\theta(\rvx_t, t)) 
\end{equation}
To train the diffusion model $\rveps_\theta(\rvx_t, t)$ that predicts the cumulative noise up to timestep $t$ for given latent $\rvx_t$, DDPM aligns the mean of the transition $p_\theta(\rvx_{t-1} | \rvx_t)$ with the posterior $q(\rvx_{t-1}| \rvx_t, \rvx_0)$: 
\begin{equation}
\label{eq:mean-align}
\min_\theta \E_{\rvx_0 \sim q_\mathrm{data}, t \sim \gU(1, T), \rveps \sim \gN(\mathbf{0},  \mathbf{I})} \| \rveps - \rveps_\theta(\sqrt{\bar{\alpha}_t} \rvx_0 + \sqrt{1-\bar{\alpha}_t}\rveps, t)  \|^2  \nonumber  \quad \text{where} \quad  \bar{\alpha}_t = \prod_{\tau=1}^t (1  - \beta_\tau)
\end{equation}
Once trained, starting from $\rvx_T \sim \gN(\mathbf{0}, \mathbf{I})$, the sampling process iteratively invokes $\rveps_\theta$:
\begin{equation}
\label{eq:sample}
\rvx_{t-1} = \rveps_\theta(\rvx_t, t),
\end{equation}
which generates the following trajectory $\{\rvx_T, \rvx_{T-1}, \ldots, \rvx_0\}$.

\subsection{Design of \system}

Prior work on training DPDMs~\citep{dockhorn2022differentially,ghalebikesabi2023differentially,lyu2023differentially} often applies DP-SGD~\citep{abadi2016deep} to fine-tune the entire sampling process using private data, resulting in significant utility loss and inference cost. However, it is known that, within a given sampling trajectory $\{\rvx_T, \rvx_{T-1}, \ldots, \rvx_0\}$, the early steps only determine the high-level image layout shared by many latents, while the later steps determine the details~\citep{sensitivity,sdedit}. Thus, instead of privatizing the end-to-end sampling process, by fully utilizing the public data, we may skip intermediate steps and focus on fine-tuning the later steps using private data.

Motivated by this idea, as illustrated in Figure~\ref{fig:NNflow}, \system first pre-trains a diffusion model $\rveps_\theta$ using the public data $\gD^\mathrm{pub}$. Further, \system builds a knowledge base $\gK\gB$ by calculating the diffusion trajectories of $\gD^\mathrm{pub}$. \system then fine-tunes $\rveps_\theta$ using the private data $\gD^\mathrm{prv}$ as follows. Corresponding to each input $\rvx \in \gD^\mathrm{prv}$, it generates its initial steps $\rvx_{T : k}$ in the sampling process, uses $\rvx_k$ (at timestep $k$) as a query to retrieve a similar trajectory $\hat{\rvx}_{T : 0}$ from $\gK\gB$, and resumes DP training the sampling process, starting from $\hat{\rvx}_v$ (at timestep $v$) of the retrieved trajectory, to reconstruct $\rvx$.
Intuitively, this RAG strategy skips the sampling process from timestep $k$ to $v$, thereby improving privacy saving, generative quality, and inference efficiency. Next, we elaborate on the implementation of \system's key components.

\subsection{Building Trajectory Knowledge Base}

We divide the public data $\gD^\mathrm{pub}$ into two parts $\gD^\mathrm{pub}_\mathrm{pre}$ and $\gD^\mathrm{pub}_\mathrm{ref}$ to avoid overfitting, with $\gD^\mathrm{pub}_\mathrm{pre}$ to pre-train the diffusion model $\rveps_\theta$ and $\gD^\mathrm{pub}_\mathrm{ref}$ to
construct the trajectory knowledge base $\gK\gB$.

\begin{algorithm}[!t]\small
\KwIn{reference data $\gD$, pre-trained diffusion model $\rveps_\theta$, timestep $k$}
\KwOut{latent feature extractor $h$}
\While{not converged yet}{
\ForEach{$\rvx \in \gD$}{
\tcp{\rm generate positive and negative pairs}
 generate $\tilde{\rvx}$, $\tilde{\rvx}^+$, and $\gN_\rvx^-$ with random augmentations\; 
\tcp{\rm generate latents at timestep $k$}
sample $\tilde{\rvx}_k$, $\tilde{\rvx}^+_k$, and $\tilde{\rvx}^-_k$ for $\tilde{\rvx}^- \in \gN_\rvx^-$ following \meq{eq:sample}\;
\tcp{\rm compute contrastive loss}
compute $\ell_\mathrm{CL}(\rvx)$ following \meq{eq:cl}\; 
\tcp{\rm update feature extractor}
update $h$ to minimize $\ell_\mathrm{CL}(\rvx)$\;
}
}
\Return $h$\;
\caption{Training latent feature extractor. \label{alg:extractor}}
\end{algorithm}

For each $\rvx \in \gD^\mathrm{pub}_\mathrm{ref}$, we construct its sampling trajectory by iteratively applying \meq{eq:sampling} to generate a sequence of latents $\{\rvx_1, \ldots, \rvx_T \}$, and store $(\rvx_k, \rvx_v)$ as a key-value pair ($k > v$) in $\gK\gB$. During RAG, 
we may sample a random latent $\rvx_T$ at timestep $T$ and generate its  early trajectory by iteratively invoking $\rveps_\theta$ (\meq{eq:sample}) until timestep $k$: $\{\rvx_T, \rvx_{T-1}, \ldots, \rvx_k\}$ and use $\rvx_k$ as a query to search for its nearest neighbors (in terms of $\ell_2$-norm) in $\gK\gB$.
For simple datasets (e.g., MNIST), due to their distributional sparsity, this straightforward approach is effective as it is possible to enforce all the trajectories to share a fixed initial latent  $\rvx_T$~\citep{zhang2023redi}. However, for more complex datasets (e.g., CIFAR10), their distributional density necessitates allowing different trajectories to have distinct 
initial latents, making this approach much less effective. More importantly, enforcing the same initial latent severely limits the model's generative quality and diversity.  

\begin{wrapfigure}{r}{0.3\textwidth}
    \centering
    \vspace{5pt}
    \includegraphics[width=0.22\textwidth]{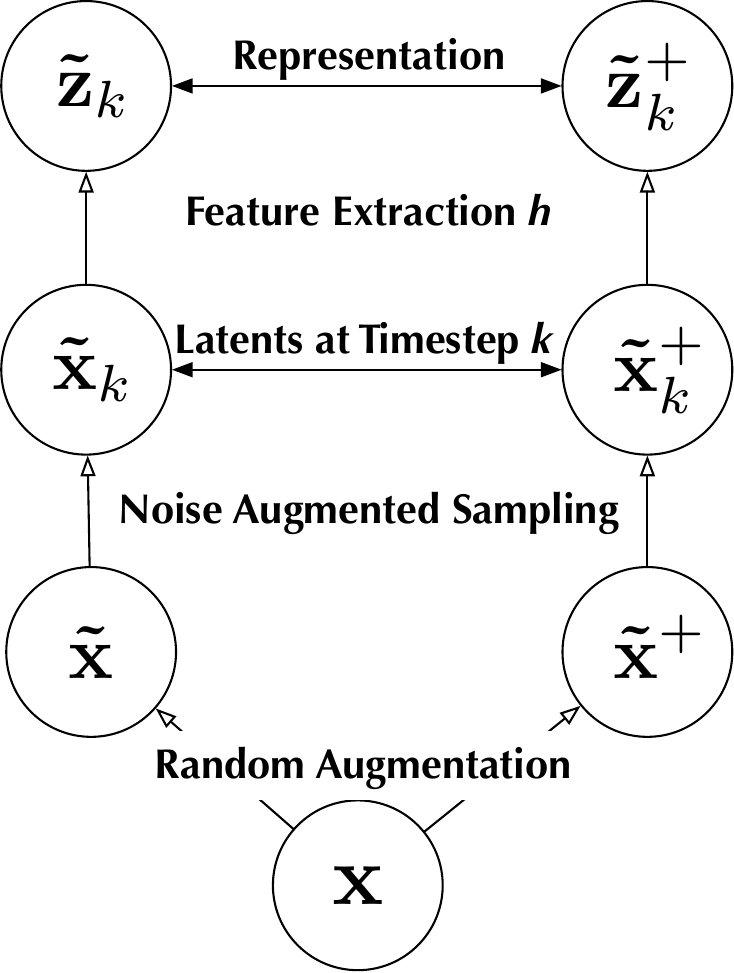}
    \caption{Contrastive learning of noise-augmented latents $\{\rvx_k\}$.}
    \label{fig:feature_extractor}
\end{wrapfigure}
Thus, instead of applying the similarity search on the latents $\{\rvx_k\}$ directly, we first extract their features (by applying a feature extractor $h$) and perform the search in their feature space. To this end, we first project $\rvx_k$ to the input space by applying one-step denoising on $\rvx_k$ using the pre-trained diffusion model $\rveps_\theta$, and then apply the feature extractor $h$ on the denoised $\rvx_k$ to extract its feature: $\rvz_k = h(\rveps_\theta(\rvx_k, k))$. For simplicity, we omit the one-step denoising in the following notations: $\rvz_k = h(\rvx_k)$.  

We employ contrastive learning~\citep{chen2020simple,chen2020improved} to train the feature extractor $h$. Intuitively, contrastive learning learns representations by aligning the features of the same input under various augmentations (e.g., random cropping) while separating the features of different inputs. In our current implementation, we extend the SimCLR~\citep{chen2020simple} framework, as illustrated in Figure~\ref{fig:feature_extractor}. Specifically, for each input $\rvx \in \gD^\mathrm{pub}_\mathrm{ref}$, a pair of its augmented views $(\tilde{\rvx}, \tilde{\rvx}^+)$ forms a ``positive'' pair, while a set of augmented views of other inputs $\gN_\rvx^-$ forms the ``negative'' samples. The contrastive loss is defined by the InfoNCE loss~\citep{oord2018representation}, which aims to maximize the similarity of positive pairs relative to that of negative pairs:
\begin{equation}
\label{eq:cl}
\ell_\mathrm{CL}(\rvx) = -  \log  \frac{\exp(\mathrm{sim}(h(\tilde{\rvx}_k), h(\tilde{\rvx}_k^+)) / \tau)}{\sum_{\tilde{\rvx}^- \in \gN_\rvx^-}  \exp(\mathrm{sim}(h(\tilde{\rvx}_k), h(\tilde{\rvx}_k^-))/\tau) + \exp(\mathrm{sim}(h(\tilde{\rvx}_k), h(\tilde{\rvx}_k^+)) / \tau)}
\end{equation}
where $\tilde{\rvx}_k$ denotes the sampled latent at timestep $k$ corresponding to $\tilde{\rvx}$ (similar for $\tilde{\rvx}_k^+$ and $\tilde{\rvx}_k^-$), $\mathrm{sim}$ is the similarity function (e.g., cosine similarity), and $\tau$ is the hyper-parameter of temperature. Algorithm~\ref{alg:extractor} sketches the training of the latent feature extractor. 

After training the latent feature extractor $h$, we build the trajectory knowledge base $\gK\gB$. For each $\rvx \in \gD^\mathrm{pub}_\mathrm{ref}$, we sample its trajectory as $(\rvx_1, \rvx_{2}, \ldots, \rvx_T)$; we consider $\rvx_k$'s feature, $\rvz_k = h(\rvx_k)$, as the key and $\rvx_v$ as the value, and store the key-value pair $(\rvz_k, \rvx_v)$ into $\gK\gB$. 

\begin{algorithm}[!t]\small
\KwIn{private data $\gD^\mathrm{prv}$, pre-trained denoiser $\rveps_\theta$, feature extractor $h$, trajectory knowledge base $\gK\gB$, batch size $B$, timestep $k$, number of iterations $I$, gradient norm bound $C$, noise scale $\sigma$}
\KwOut{fine-tuned diffusion model $\rveps_\theta$}
\For{$i \in [I]$}{
sample a batch $\gB$ of size $B$ from $\gD^\mathrm{prv}$ via Poisson sampling\;
\ForEach{$\rvx \in \gB$}{
\tcp{\rm find nearest neighbor}
sample $\rvx_k$ following \meq{eq:sample}\;
find key $\hat{\rvz}_k$ closest to $h(\rvx_k)$ in $\gK\gB$\;
\tcp{\rm skip intermediate steps}
fetch value $\hat{\rvx}_v$ corresponding to $\hat{\rvz}_k$\;
compute gradient $\rvg(\rvx) \gets \nabla_\theta \ell_\mathrm{DM}(\rvx, \hat{\rvx}_v)$ following \meq{eq:diffloss}\;
\tcp{\rm clip gradient}
$\tilde{\rvg}(\rvx) \gets \rvg(\rvx) / \max(1, \frac{\|\rvg(\rvx)\|}{C})$\;
}
\tcp{\rm apply DP noise}
$\tilde{\rvg}(\gB) \gets \frac{1}{B} \sum_{\rvx \in \gB} \tilde{\rvg}(\rvx) + \frac{C}{B} \gN(\mathbf{0}, \sigma^2 \mathbf{I})$ \;
$\theta \gets \mathrm{Adam}(\theta,  \rev{\tilde{\rvg}}(\gB))$\;
}
\Return $\rveps_\theta$\;
\caption{\system. \label{alg:rapid}}
\end{algorithm}

\subsection{Training Differentially Private Diffusion Model}

Leveraging the trajectory knowledge base $\gK\gB$, we further train the denoiser $\rveps_\theta$ on private data $\gD^\mathrm{prv}$. Notably, we focus on training $\rveps_\theta$ from timestep $v$ to $0$, leading to the advantages of fully utilizing limited private data and reducing overall privacy costs.

As outlined in Algorithm~\ref{alg:rapid}, at each iteration, we sample batch $\gB$ from $\gD^\mathrm{prv}$ using Poisson sampling for privacy amplification~\citep{renyi-gm}. For each input $\rvx \in \gB$, we \mct{i} sample its early trajectory $\rvx_k$ up to timestep $k$, \mct{ii} use its feature $\rvz_k = h(\rvx_k)$ as a query to find $\rvz_k$'s nearest neighbor $\hat{\rvz}_k$ in $\gK\gB$, \mct{iii}
reuse the value $\hat{\rvx}_v$ corresponding to $\hat{\rvz}_k$ in $\gK\gB$ as the starting point at timestep $v$, and \mct{iv} train $\rveps_\theta$ to reconstruct $\rvx$. In other words, $\rveps_\theta$ is fine-tuned to predict the random noise $(\rvx - \hat{\rvx}_v)$ at timestep $v$. 
To make the training differentially private, we extend DP-SGD~\citep{abadi2016deep} during updating $\rveps_\theta$. For each input $\rvx$ in a batch $\gB$, we compute its diffusion loss as: 
\begin{equation}
\label{eq:diffloss}
\ell_\mathrm{DM}(\rvx, \hat{\rvx}_v) = \sE_{v' \sim \gU(1, v) } \| \frac{\hat{\rvx}_v -\sqrt{\bar{\alpha}_v}\rvx}{\sqrt{1-\bar{\alpha}_v}} - \rveps_\theta( 
\sqrt{\bar{\alpha}_{v'}}\rvx + \frac{\sqrt{1-\bar{\alpha}_{v'}}}{\sqrt{1-\bar{\alpha}_{v}}} (\hat{\rvx}_v -\sqrt{\bar{\alpha}_v}\rvx)
, v')  \|
\end{equation}
where the random noise $(\rvx - \hat{\rvx}_v)$ is scaled with a randomly sampled timestep $v'$. We compute the  gradient $\rvg(\rvx) = \nabla_\theta \ell_\mathrm{DM}(\rvx, \hat{\rvx}_v)$. To bound $\rvg(\rvx)$'s influence on $\rveps_\theta$, we clip 
$\rvg(\rvx)$ using its $\ell_2$ norm. We then sanitized the per-batch gradient as $\tilde{\rvg}(\gB)$ by applying random Gaussian noise before updating $\rveps_\theta$ using the Adam optimizer~\citep{adam}. 

We prove the privacy guarantee of Algorithm~\ref{alg:rapid} under R\'{e}nyi differential privacy (RDP)~\citep{renyi}, which can be converted to $(\epsilon, \delta)$-DP. The following theorem formulates the guarantee provided by \system (proof deferred to \msec{sec:proof}).  
\begin{theorem}
\label{the:main}
Using the sanitized per-batch gradient 
$\tilde{\rvg}(\gB)$ to update $\rveps_\theta$ satisfies $(\alpha, \frac{2\alpha}{ \sigma^2})$-RDP.
\end{theorem}
The overall privacy cost of \system is computed via RDP composition~\citep{renyi}, which can be further improved using more advanced privacy accounting~\citep{privacy-accounting}.

The inference of \system runs as follows. By sampling random Gaussian noise $\rvx_T$ at timestep $T$, we generate its early trajectory $\rvx_k$ up to timestep $k$; using the feature $\rvz_k$ of $\rvx_k$ as the query, we search for $\rvz_k$'s nearest neighbor $\hat{\rvz}_k$ in $\gK\gB$; we then use the corresponding value $\hat{\rvx}_v$ as the starting point at timestep $v$ to resume the sampling. Compared with prior work~\citep{lyu2023differentially,dockhorn2022differentially}, this RAG-based inference also significantly improves inference efficiency.



%% file: evaluation.tex
\section{Evaluation}



\subsection{Experimental Setting}

{\bf Datasets.} We focus on the image synthesis task. In each task, we use the public dataset $\gD^\mathrm{pub}$ to pre-train the diffusion model and build the trajectory knowledge base for the retrieval-augmented generation, and use the private dataset $\gD^\mathrm{pub}$ to further fine-tune/train the diffusion model in a differentially private manner. Specifically, we consider the following 4 settings: i) EMNIST~\citep{cohen2017emnist} (public) and MNIST~\citep{deng2012mnist} (private), ii) ImageNet32~\citet{deng2009imagenet} (public) and CIFAR10~\citep{krizhevsky2009learning} (private), iii)  FFHQ32~\citep{karras2019style} (public) and CelebA32~\citep{liu2015faceattributes} (private), and iv) FFHQ64~\citep{karras2019style} (public) and CelebA64~\citep{liu2015faceattributes} (private). More details of these datasets are deferred to Table~\ref{table:setting}.


{\bf Diffusion models.} We primarily use the latent diffusion model~\citep{rombach2022high} as the underlying diffusion model and DDIM~\citep{song2020denoising} as the default sampler.

{\bf Baselines.} We mainly consider two state-of-the-art DPDM methods as baselines: differential private diffusion model (\dpdm)~\citep{dockhorn2022differentially} and differential private latent diffusion model (\dpldm)~\citet{lyu2023differentially}.

{\bf Metrics.} Following prior work~\citep{dockhorn2022differentially,lyu2023differentially}, we use the Frechet Inception Distance (FID) score to measure the generative quality of different methods. In addition, we adopt the coverage metric~\citep{naeem2020reliable} to measure the generative diversity. Intuitively, given a reference real dataset $\gD^\mathrm{real}$, the coverage is measured by the proportion of samples from $\gD^\mathrm{real}$ that have at least one sample from the synthesized data $\gD^\mathrm{syn}$ in their neighborhood (with neighborhood size fixed as 5~\citep{lebensold2024dp}). Formally, 
\begin{equation}
\text{Coverage} = \frac{1}{|\gD^\mathrm{real}|} \sum_{\rvx \in \gD^\mathrm{real}} \bm{1}_{\exists \rvx' \in \gD^\mathrm{syn} \wedge \rvx' \in \gN_\rvx}
\end{equation}
where $\bm{1}$ is the indicator function and $\gN_\rvx$ denotes $\rvx$'s neighborhood.

{\bf Privacy.} We use {\sc Opacus}~\citet{opacus}, a DP-SGD library, for DP training and privacy accounting. Following prior work~\citep{dockhorn2022differentially}, we fix the setting of $\delta$ as $10^{-5}$ for the CIFAR10 and MNIST datasets and $10^{-6}$ for CelebA dataset so that $\delta$ is smaller than the reciprocal of the number of training samples. Similar to existing work, we also do not account for the (small) privacy cost of hyper-parameter tuning.

To simulate settings with modest compute resources, all the experiments are performed on a workstation running one Nvidia RTX 6000 GPU. 

\subsection{Main Results}

We empirically evaluate \system and baselines. To make a fair comparison, we fix the default batch size as 64 for \system and \dpldm; we do not modify the batch size (i.e., 8,192) for \dpdm because the impact of batch size on its performance is so significant that it stops generating any recognizable images with smaller batch sizes. By default, we fix the sampling timesteps as 100 across all the methods. For MNIST, we train the diffusion model under three privacy settings $\epsilon = \{0.2, 1, 10\}$, corresponding to the low, medium, and high privacy budgets; for the other datasets, we vary the privacy budget as $\epsilon = \{1, 10\}$.


\begin{table}[!ht]\small
\renewcommand{\arraystretch}{1.2}
\centering
\begin{tabular}{c|c|c|c|c}
Setting & Privacy ($\epsilon)$ & \dpdm & \dpldm & \system  \\
\hline
\multirow{3}{*}{EMNIST$\rightarrow$MNIST} & 0.2 & 125.7 & 50.8  & \cellcolor{Red}24.0 \\
& 1  & 50.5 & 34.9  & \cellcolor{Red}18.5 \\
& 10  & \cellcolor{Red}12.9 & 27.2  & 14.1 \\ 
\hline
\multirow{2}{*}{ImageNet32$\rightarrow$CIFAR10} & 1 & $\backslash$ & 79.1  & \cellcolor{Red}63.2 \\
& 10 & 109.9 & 33.3 & \cellcolor{Red}25.4 \\
\end{tabular}
\caption{FID scores of class-conditional generation by different methods. \label{tab:cond}}
\end{table}


{\bf Class-conditional generation.} We evaluate the quality of class-conditional generation by different methods, in which, besides the input image $\rvx$, a guidance signal $y$ (e.g., $\rvx$'s class label) is also provided for training (and inference). We consider the EMNIST$\rightarrow$MNIST and ImageNet32$\rightarrow$CIFAR10 settings. Table~\ref{tab:cond} compares the generative quality of different methods. 
Observe that, under the same privacy budget, \system considerably outperforms the baselines across most cases. For instance, under the ImageNet32$\rightarrow$CIFAR10 setting, with $\epsilon =1$, \system attains an FID score of 63.2, while \dpdm fails to produce any sensible outputs, highlighting the effectiveness of RAG in facilitating the DP training of diffusion models. 


\begin{table}[!ht]\small
\renewcommand{\arraystretch}{1.2}
\centering
\begin{tabular}{c|c|c|c|c}
Setting & Privacy ($\epsilon)$ & \dpdm & \dpldm  & \system \\
\hline
\multirow{3}{*}{EMNIST$\rightarrow$MNIST (CNN)}  & 0.2  & 85.77$\%$ &  11.35$\%$ & \cellcolor{Red}96.43$\%$\\
& 1  & 95.18$\%$ & 74.62$\%$  & \cellcolor{Red}98.11$\%$\\
& 10  & 98.06$\%$ & 95.54$\%$  & \cellcolor{Red}99.04$\%$\\ 
\hline
\multirow{2}{*}{ImageNet32$\rightarrow$CIFAR10 (ResNet)} & 1 & $\backslash$ & 50.39$\%$ & \cellcolor{Red}63.61$\%$\\
& 10 & 30.41$\%$ & 66.02$\%$  & \cellcolor{Red}67.37$\%$\\
\end{tabular}
\caption{Downstream accuracy of classifiers trained on synthesized data. \label{tab:Downstream}}
\end{table}


We further evaluate the utility of the data synthesized by different methods to train downstream classifiers. For a fair comparison, we use the same classifier architecture to measure the downstream accuracy. For the synthesized MNIST data, we train a Convolutional Neural Network (CNN)~\citep{krizhevsky2012imagenet} and test its performance on the MNIST testing set. For the synthesized CIFAR10 data, we train a ResNet-9~\citep{he2016deep} and evaluate its accuracy on the CIFAR10 testing set, with results summarized in Table~\ref{tab:Downstream}. Observe that the classifier trained on \system's synthesized data
largely outperforms the other methods. For instance, under the ImageNet32$\rightarrow$CIFAR10 setting with $\epsilon=1$, \system attains 63.6\% downstream accuracy with 10\% higher than the baselines, indicating the high utility of the data synthesized by \system.

\begin{figure}[!ht]
    \centering
    \includegraphics[width=0.9\linewidth]{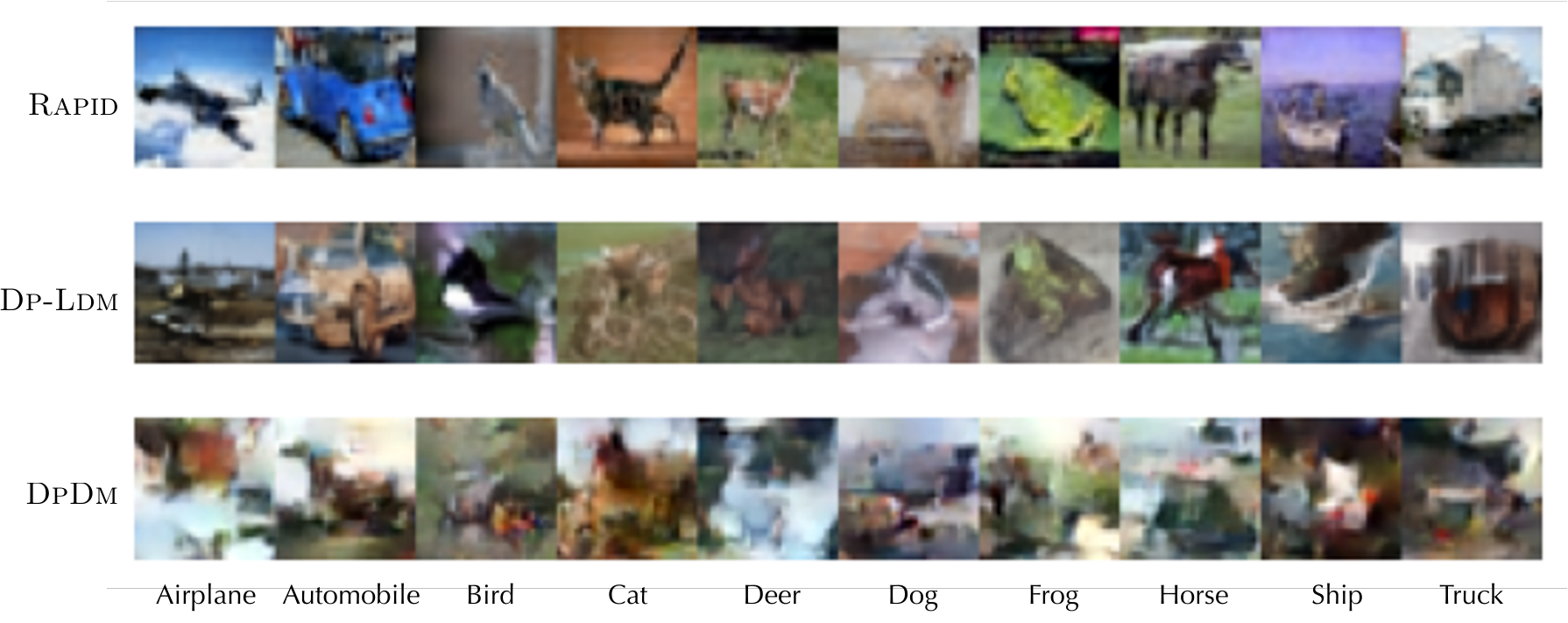}
    \caption{Random samples synthesized by \system and baselines trained under the ImageNet32$\rightarrow$CIFAR10 setting with $\epsilon = 10$.
    \label{fig:Cifar10}}
\end{figure}

We also qualitatively compare the class-conditional samples generated by \dpdm, \dpldm, and \system trained under the ImageNet32$\rightarrow$CIFAR10 setting (with $\epsilon = 10$), as shown in Figure\mref{fig:Cifar10}. It is observed that across different classes, \system tends to produce samples of higher visual quality, compared with the baselines. 


\begin{table}[!t]\small
\renewcommand{\arraystretch}{1.2}
\centering
\begin{tabular}{c|c|c|c|c}
Setting & Privacy ($\epsilon)$ & \dpdm & \dpldm &  \system \\
\hline
\multirow{2}{*}{FFHQ32$\rightarrow$CelebA32} & 1  & 135.9 | 0.087 & 65.3 | 0.74 & \cellcolor{Red}52.8 | 0.96\\
& 10  & 29.8 | 0.55 & 38.0 | 0.98  & \cellcolor{Red}28.0 | 0.98\\
\hline
\multirow{2}{*}{FFHQ64$\rightarrow$CelebA64} & 1 & $\backslash$ &  72.2 | 0.59    & \cellcolor{Red}60.5 | 0.90 \\ 
& 10 & 80.8 | 0.094 & 45.2 | 0.94 & \cellcolor{Red}37.3 | 0.93 \\ 
\end{tabular}
\caption{FID (left) and coverage (right) scores of unconditional generation by different methods. \label{tab:uncond}}
\end{table}


{\bf Unconditional generation.} We further evaluate the unconditional generation by different methods under the FFHQ32$\rightarrow$CelebA32 and FFHQ64$\rightarrow$CelebA64 settings. Table~\ref{tab:uncond} summarizes the FID and coverage scores of different methods. Observe that \system outperforms the baselines by large margins in terms of generative quality (measured by the FID score). For instance, in the case of FFHQ32$\rightarrow$CelebA32 under $\epsilon = 10$, \system achieves an FID score of 37.3, which is 19.1\% and 51.8\% lower than \dpldm and \dpdm, respectively, highlighting its superior generative quality. Meanwhile, across all the cases, \system attains the highest (or the second highest) generative diversity (measured by the coverage score). Overall, \system strikes the optimal balance between generative quality and diversity among all three methods.


\begin{figure}[htp]
    \centering
    \includegraphics[width=0.9\textwidth]{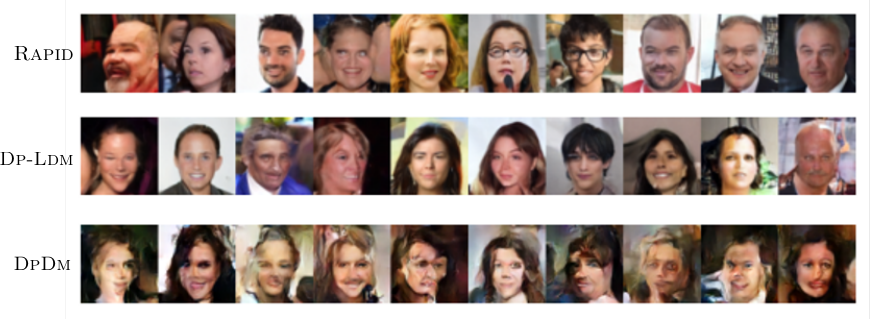}
    \caption{Random samples synthesized by \system and baselines trained under the setting of FFHQ64$\rightarrow$CelebA64 (with $\epsilon = 10$).}
    \label{fig:CelebA64}
\end{figure}

We also qualitatively compare the unconditional samples generated by different methods trained on the CelebA64 dataset (with $\epsilon = 10$). Figure\mref{fig:Cifar10} shows random samples synthesized by \dpdm, \dpldm, and \system (more samples in \msec{sec:additional}). Observe that in general \system tends to produce unconditional samples of higher visual quality, compared with the baselines.

\subsection{Ablation Studies}
\label{sec:ablation}

Next, we conduct ablation studies to understand the impact of various key factors, such as batch size and knowledge base size, on \system's performance. We use the FFHQ32$\rightarrow$CelebA32 (with $\epsilon = 10$) 
as the default setting.

{\bf Retrieval accuracy.} Recall \system relies on retrieving the most similar trajectory from the knowledge base. A crucial question is thus whether \system indeed retrieves semantically relevant neighbors. To answer this question, under the class-conditional generation, we evaluate the accuracy of \system in retrieving the neighbors from the class corresponding to the given label (e.g., ``automobile''). 
We calculate the top-$k$ ($k = 1, 5$) accuracy based on the true labels of the retrieved neighbors. Under the setting of ImageNet32$\rightarrow$CIFAR10, \system attains 81.2$\%$ top-1 accuracy and 94.4$\%$ top-5 accuracy, respectively, indicating its effectiveness. 




\begin{wrapfigure}{r}{0.5\textwidth}
    \centering
   \vspace{5pt}
\includegraphics[width=0.49\textwidth]{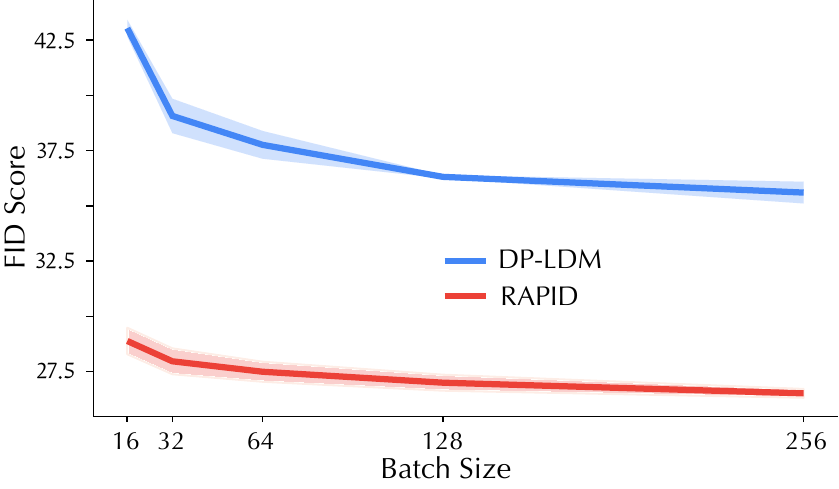}
    \caption{Impact of batch size.}
    \label{fig:batchsize}
\end{wrapfigure}
\textbf{Batch size.} In contrast to existing methods (e.g., \dpdm and \dpldm) that typically require excessively large batch sizes (e.g., 8,192 samples per batch), by fully utilizing the public data using its RAG design, \system can generate high-quality samples under small batch sizes. Here, we evaluate the impact of batch-size setting on the performance of \system and \dpldm, with results illustrated in Figure~\ref{fig:batchsize}. It can be noticed that \system attains an FID score of 29.5 under a batch size of 16 while its score steadily improves to 26.67 as the batch size increases from 16 to 256, highlighting its superior performance under small batch sizes. 

\begin{table}[!ht]\small
\renewcommand{\arraystretch}{1.2}
\centering
\begin{tabular}{c|c|c|c|c}
Setting & Privacy ($\epsilon)$ & \dpdm & \dpldm & \system \\
\hline
\multirow{2}{*}{FFHQ32$\rightarrow$CelebA32} & 1  & 153.1 & 72.2 & \cellcolor{Red}56.6\\
& 10  & 33.0 & 42.6 & \cellcolor{Red}30.3 \\
\hline
\multirow{2}{*}{FFHQ64$\rightarrow$CelebA64} & 1 & $\backslash$ &  78.7  & \cellcolor{Red}68.9 \\ 
& 10 & 86.2 & 50.3 & \cellcolor{Red}41.1 \\ 
\end{tabular}
\caption{FID scores of unconditional generation with the PNDM sampler. \label{tab:pbdm}}
\end{table}

{\bf Sampler.} By default, we use DDIM~\citep{song2020denoising} as the underlying sampler. Here, we evaluate the influence of the sampler on \system. Specifically, we consider the state-of-the-art PNDM sampler~\citep{liu2022pseudo} and evaluate the performance of different methods in unconditional generation tasks. By comparing Table~\ref{tab:uncond} and Table~\ref{tab:pbdm}, it is observed that \system achieves similar FID scores in both cases, indicating its insensitivity to the underlying sampler.

\begin{wrapfigure}{r}{0.5\textwidth}
    \centering
   \vspace{5pt}
\includegraphics[width=0.49\textwidth]{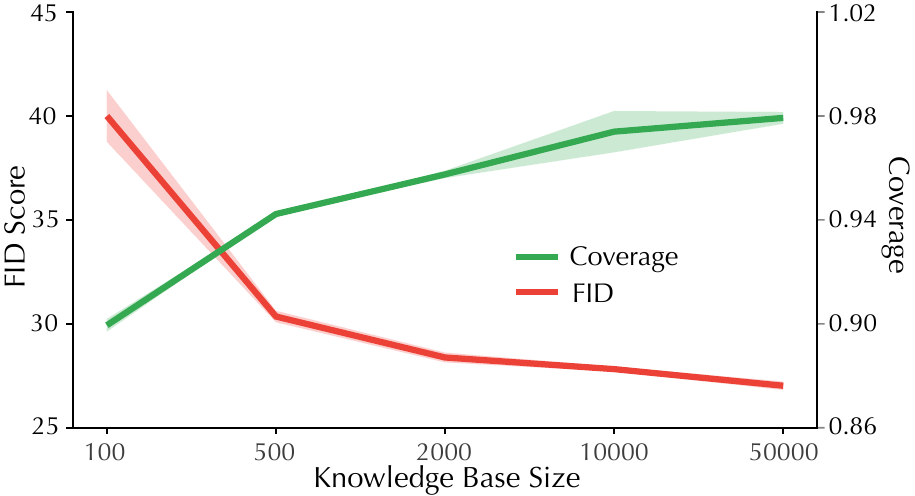}
    \caption{Impact of knowledge-base size.}
    \label{fig:kb}
\end{wrapfigure}
\textbf{Knowledge base size.} One key component of \system is its trajectory knowledge base that supports RAG. We now evaluate the impact of the knowledge base size. As shown in Figure~\ref{fig:kb}, we evaluate how \system's performance varies as the knowledge-base size grows from 100 to 50,000. As expected, both \system's FID and coverage scores improve greatly with the knowledge-base size. Meanwhile, even under a small knowledge base (e.g., of size 100), \system attains satisfactory performance (e.g., with an FID score of about 40 and a coverage of about 0.9).

\begin{wrapfigure}{r}{0.5\textwidth}
    \centering
   \vspace{-25pt}
\includegraphics[width=0.49\textwidth]{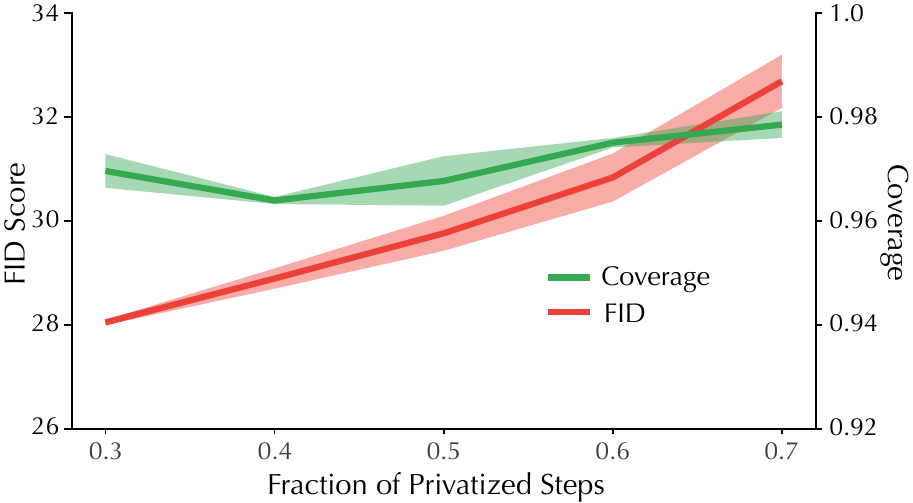}
    \caption{Impact of the fraction of privatized steps.}
    \label{fig:steps}
\end{wrapfigure}
\textbf{Fraction of privatized steps.} Recall that, over the sample trajectory, \system samples the initial $(T - k)$ steps, skips the intermediate $(k - v)$ steps, and privatizes the later $v$ steps. By default, we set $k/T$ = 0.8 and $v/T$ = 0.2. We now evaluate the influence of the fraction of privatized steps $v/T$ on \system.
Figure~\ref{fig:steps} shows how \system's FID score varies as $v/T$ increases from 0.3 to 0.7 (with $k/T$ fixed as 0.8) under the FFHQ32$\rightarrow$CelebA32 setting (with $\epsilon = 10$). Notably, as more steps are privatized, \system's FID score deteriorates while its coverage score improves marginally, suggesting an interesting trade-off between the generative quality and diversity. Intuitively, a larger fraction of $v/T$ indicates less reliance on the retrieved trajectory, encouraging more diverse generations but negatively impacting the generative quality.





%% file: conclusion.tex
\section{Conclusion and Future Work}

The work represents a pilot study of integrating retrieval-augmented generation (RAG) into the private training of generative models. We present \system, a novel approach for training differentially private (DP) diffusion models. Through extensive evaluation using benchmark datasets and models, we demonstrate that \system largely outperforms state-of-the-art methods in terms of generative quality, memory footprint, and inference efficiency. The findings suggest that integrating RAG with DP training represents a promising direction for designing privacy-preserving generative models.

This work also opens up several avenues for future research. \mct{i}
Like other approaches for training DP diffusion models and the broader pre-training/fine-tuning paradigm, \system relies on access to a diverse public dataset that captures a range of patterns and shares similar high-level layouts with the private data. It is worth exploring scenarios with highly dissimilar public/private data~\citep{liu2021leveraging,liu2021iterative,fuentes2024joint}.
\mct{ii} In its current implementation, \system retrieves only the top-1 nearest trajectory in RAG. Exploring ways to effectively aggregate multiple neighboring trajectories could improve generative quality and diversity. \mct{iii} While \system's privacy accounting focuses on privatizing the fine-tuning stage, it is worth accounting for random noise introduced by the diffusion process to further improve its privacy guarantee~\citep{wangdp}. \mct{iv} Although this work primarily focuses on image synthesis tasks, given the increasingly widespread use of diffusion models, extending \system to other tasks (e.g., text-to-video generation) presents an intriguing opportunity. 



%% file: appendix.tex
\appendix

\section{Proofs}
\label{sec:proof}

\begin{definition}{\rm (R\'{e}nyi differential privacy~\citep{renyi})} A randomized mechanism $\gM : \gD \rightarrow \gR$ over domain $\gD$ and
range $\gR$ satisfies $(\alpha, \epsilon)$-RDP if for any two adjacent $d, d' \in D$: $\gD_\alpha(\gM(d)|\gM(d')) \leq \epsilon$, 
where $D_\alpha$ denotes the R\'{e}nyi divergence of order $\alpha$.
\end{definition}

RDP can be converted to DP. If an mechanism satisfies $(\alpha, \epsilon)$-RDP, it also satisfies $(\epsilon + \frac{\log 1/\delta}{\alpha - 1}, \delta)$-DP~\citep{renyi}.

Notably, Gaussian mechanism can provide RDP. Specifically, for function $f$ with sensitivity $\Delta f = \max_{d,d'} \|f(d)-f(d') \|_2$, releasing $f(d) + \gN(0, \sigma^2)$ satisfies $(\alpha, \frac{\alpha \Delta f}{2\sigma^2})$-RDP~\citep{renyi-gm}.

Now, we prove Theorem~\ref{the:main}. Recall that Algorithm~\ref{alg:rapid} computes the per-sample gradient $\rvg(\rvx)$ and clips it to bound its influence $\tilde{\rvg}(\rvx) \gets \rvg(\rvx) / \max(1, \frac{\|\rvg(\rvx)\|}{C})$. It then computes the per-batch gradient 
$\rvg(\gB) \leftarrow \frac{1}{B} \sum_{\rvx \in \gB} \tilde{\rvg}(\rvx)$ and applies Gaussian noise:
$\tilde{\rvg}(\gB) \gets \rvg(\gB)  + \frac{C}{B} \gN(\mathbf{0}, \sigma^2 \mathbf{I})$.
\begin{proof}
Consider two adjacent mini-batches $\gB$ and $\gB'$ that differ by one sample $\rvx^-/\rvx^+$:
$ \gB' = \gB  \setminus  \{\rvx^-\} \cup \{\rvx^+\}$. We bound the difference of their batch-level gradients as follows: 
\begin{equation}
\begin{split}
\| \rvg(\gB)  - \rvg(\gB')  \|_2  & =  
\| \frac{1}{B} \sum_{\rvx \in \gB} \tilde{\rvg}(\rvx) - \frac{1}{B} \sum_{\rvx \in \gB'} \tilde{\rvg}(\rvx)\|_2
\\
& = \| \frac{1}{B}\tilde{g}(\rvx^-) -  \frac{1}{B}\tilde{g}(\rvx^+) \|_2 \\
& = \frac{1}{B} \sqrt{ 
\| \tilde{g}(\rvx^-) \|^2_2  + \| \tilde{g}(\rvx^+) \|^2_2 - 2 \tilde{g}^T(\rvx^-) \tilde{g}(\rvx^+)
}\\
& \leq \frac{1}{B} \sqrt{C^2 + C^2 + 2C^2} \\
& = \frac{2C}{B}
\end{split}
\end{equation}
where we use the fact that $\tilde{g}(\rvx^-)$
and $\tilde{g}(\rvx^+)$ are bounded by $C$ and the Cauchy-Schwarz inequality.

Thus, the sensitivity of $\rg(\gB)$ is $\frac{2C}{B}$. Following the RDP Gaussian mechanism, releasing the sanitized batch-level gradient $\tilde{g}(\gB)$ provides $(\alpha, \frac{2\alpha}{\sigma^2})$-RDP, corresponding to 
$(\frac{2\alpha}{\sigma^2} + \frac{\log 1/\delta}{ \alpha -1}, \delta)$-DP.
\end{proof}

\section{Experimental Setting}
\label{sec:setting}

Table~\ref{table:setting} summarizes the setting of public and private datasets in our experiments.

\begin{table}[ht!]\small
\centering
\renewcommand{\arraystretch}{1.2}
\begin{tabular}{c|c|l}
\multicolumn{2}{c|}{Public dataset $\gD^\mathrm{pub}$} & \multirow{2}{*}{Private dataset $\gD^\mathrm{prv}$} \\
\cline{1-2}
Pre-training ($\gD^\mathrm{pub}_\mathrm{pre}$ & Trajectory knowledge base $\gD^\mathrm{pub}_\mathrm{ref}$ &  \\
 \hline
 EMNIST (50K) & EMNIST (10K) &  MNIST \\
 ImageNet32 (1.2M)  &  ImageNet32 (70K)~\citep{darlow2018cinic}   & CIFAR10  \\
 FFHQ32 (60K)&FFHQ32 (10K)  & CelebA32\\
 FFHQ64 (60K)& FFHQ64 (10K) & CelebA64 \\
\end{tabular}
\caption{Setting of public/private datasets in experiments. \label{table:setting}}
\end{table}

Table~\ref{tab:autoencoder} lists the default parameter setting for training the autoencoder in the latent diffusion model under different settings, while Table~\ref{tab:pub_diff} summarizes the default parameter setting for training the diffusion model.

\begin{table}\small
\centering
\renewcommand{\arraystretch}{1.2}
\setlength{\tabcolsep}{1pt}
\begin{tabular}{r|c|c|c|c}
& EMNIST$\rightarrow$MNIST & ImageNet$\rightarrow$CIFAR10 & FFHQ32$\rightarrow$CelebA32 & FFHQ64$\rightarrow$CelebA64 \\ \hline
Input size & 32$\times$32$\times$3 & 32$\times$32$\times$3 & 32$\times$32$\times$3 & 64$\times$64$\times$3 \\ 
$z$-shape & $4 \times 4 \times 3$ & $16 \times 16 \times 3$ & $16 \times 16 \times 3$ & $64 \times 64 \times 3$ \\ 
Channels & 128 & 128 & 128 & 192 \\ 
Channel multiplier & $[1,2,3,5]$ & $[1,2]$ & $[1,2]$ & $[1,2]$ \\ 
Attention resolutions & $[32,16,8]$ & $[16,8]$ & $[16,8]$ & $[16,8]$ \\
\# ResBlocks & 2 & 2 & 2 & 2 \\ 
Batch size & 64 & 64 & 64 & 64 \\ 
\end{tabular}
\caption{Hyper-parameters for training autoencoders under different settings.}
\label{tab:autoencoder}
\end{table}

\begin{table}\small
\centering
\renewcommand{\arraystretch}{1.2}
\setlength{\tabcolsep}{1pt}
\begin{tabular}{r|c|c|c|c}
& EMNIST$\rightarrow$MNIST & ImageNet$\rightarrow$CIFAR10 & FFHQ32$\rightarrow$CelebA32 & FFHQ64$\rightarrow$CelebA64 \\ \hline
Input size & 32$\times$32$\times$3 & 32$\times$32$\times$3 & 32$\times$32$\times$3 & 64$\times$64$\times$3 \\ 
$z$-shape & $4 \times 4 \times 3$ & $16 \times 16 \times 3$ & $16 \times 16 \times 3$ & $32 \times 32 \times 3$ \\ 
\# Channels & 64 & 128 & 192 & 192 \\ 
Channel multiplier & $[1,2]$ & $[1,2,2,4]$ & $[1,2,4]$ & $[1,2,4]$ \\ 
Attention resolutions & $[1,2]$ & $[1,2,4]$ & $[1,2,4]$ & $[1,2,4]$ \\ 
\# ResBlocks & 1 & 2 & 2 & 2 \\ 
\# Heads & 2 & 8 & 8 & 8 \\ 
Batch size & 64 & 64 & 64 & 32 \\ 
Spatial transformer & True & True & False & False \\ 
Cond\_stage\_key & class\_label & class\_label & class\_label & class\_label \\ 
Conditioning\_key & crossattn & crossattn & crossattn & crossattn \\ 
\# Classes & 26 & 1000 & 1000 & 1000 \\ 
Embedding dimensions & 5 & 512 & 512 & 512 \\ 
Transformer depth & 1 & 2 & 2 & 2 \\ 
\end{tabular}
\caption{Hyper-parameters for diffusion models under different settings.}
\label{tab:pub_diff}
\end{table}

\section{Additional Results}
\label{sec:additional}

\subsection{Qualitative Comparison}

\begin{figure}
    \centering
    \includegraphics[width=0.95\textwidth]{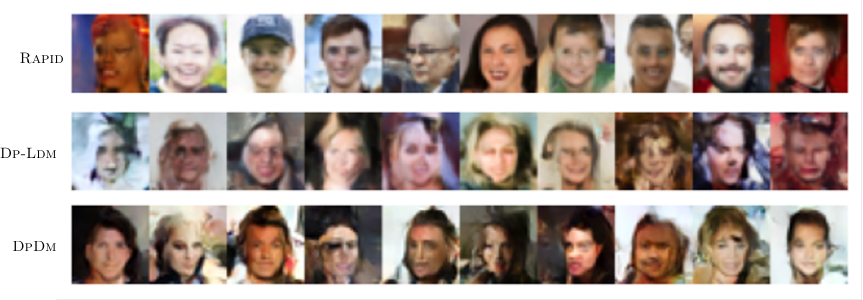}
    \caption{Random samples synthesized by \system and baselines trained under the FFHQ32$\rightarrow$CelebA32 setting (with $\epsilon = 10$).}
    \label{fig:CelebA32}
\end{figure}

Figure~\ref{fig:CelebA32} illustrates random samples synthesized by \system and baselines under the setting of FFHQ32$\rightarrow$CelebA32 (with $\epsilon = 10$).

\subsection{Dissimilar Public/Private Data}

To evaluate \system's robustness to the distributional shift between public and private data, we conduct additional experiments to evaluate \system's performance when using dissimilar public/private datasets. Specifically, we use ImageNet32 as the public dataset with added Gaussian noise $\mathcal{N}(0, 0.1)$ to degrade its quality. For the private dataset, we use VOC2005~\citep{voc} (resized to 32$\times$32), a dataset used for object detection challenges in 2005, which significantly differs from ImageNet32 and contains only about 1K images. We apply \system in this challenging setting, with results shown in Table~\ref{tab:voc}. Notably, \system outperforms baselines (e.g., \dpldm) in terms of FID scores across varying $\epsilon$, indicating its robustness to dissimilar public/private datasets.

\begin{table}[!ht]\small
\renewcommand{\arraystretch}{1.2}
\centering
\begin{tabular}{c|c|c|c|c}
Privacy ($\epsilon)$ & \dpldm & \dpsda & {\sc PrivImage} & \system \\
\hline
1  &  164.85 & 142.20 & 139.07 & 93.17 \\
10  & 147.86 & 130.42 & 123.89  & 82.56  \\
\end{tabular}
\caption{Performance of \system and baselines (measured by FID scores) in the ImageNet32$\rightarrow$VOC2005 case. \label{tab:voc}}
\end{table}

Moreover, we compare \system's performance (without DP) to direct training on the VOC2005 dataset. \system improves the FID score from 77.83 to 54.60, highlighting its ability to effectively leverage the public data even when it differs substantially from the private data.

\subsection{Additional Baselines}

We further compare \system with more recent work on DP diffusion models.
\dpsda~\citep{dpsda} synthesizes a dataset similar to the private data by iteratively querying commercial image generation APIs (e.g., DALL-E 2) in a DP manner. For fair comparison with \system, instead of using commercial APIs trained on vast datasets (hundreds of millions of images), following the setting of~\cite{privimage} that replicates \dpsda's results, we use ImageNet32 for pre-training the public model (also as the query API for \dpsda) and CIFAR10 as the private dataset.

\begin{table}[!ht]\small
\renewcommand{\arraystretch}{1.2}
\centering

\begin{tabular}{c|c|c|c|c}
\multirow{2}{*}{Privacy ($\epsilon)$} & \multicolumn{2}{c|}{Model Size = 90M} & \multicolumn{2}{c}{Model Size = 337M} \\
\cline{2-5}
& \dpsda & \system & \dpsda & \system \\
\hline
1  &  113.6 & 63.2 &  89.1 &   66.5  \\
10  & 60.9 & 25.4 & 43.8  & 29.0 \\
\end{tabular}
\caption{Performance of \dpsda and \system (measured by FID scores) in the ImageNet32$\rightarrow$CIFAR10 case. \label{tab:dpsda}}
\end{table}

Note \system and \dpsda represent two distinct approaches to training DP diffusion models, with the pre-trained model size affecting their performance differently.

For \dpsda, which uses DP evolution rather than DP training to synthesize data, larger pre-trained models tend to lead to better performance. This is demonstrated in \dpsda's ablation study~\citep{dpsda}, where increasing the model size from 100M to 270M parameters improves results by enhancing the quality of selected data.
In contrast, methods involving DP training (such as \dpdm, \dpldm, {\sc PrivImage}, and \system) may not benefit from heavily over-parameterized models, as shown in~\citep{dockhorn2022differentially}. This is because the $\ell_2$-norm noise added in DP-SGD typically grows linearly with the number of parameters.

To empirically evaluate how model complexity affects different approaches, we conduct experiments varying the size of the pre-trained model from 90M to 337M parameters (by increasing the latent diffusion model's architecture from 128 to 192 channels and expanding its residual blocks from 2 to 4).
Table~\ref{tab:dpsda} compares the performance (measured by FID scores) of \dpsda and \system across different pre-trained model sizes. As model complexity increases, \dpsda achieves better FID scores, while \system shows only marginal performance degradation. Notably, when using the same public dataset and pre-trained model, \system consistently outperforms \dpsda, suggesting that it is more effective at leveraging public data under DP constraints.

{\sc PrivImage}~\citep{privimage} uses the fine-tuning approach, querying the private data distribution to select semantically similar public samples for pretraining, followed by DP-SGD fine-tuning on the private data. The table below compares \system and {\sc PrivImage}'s performance across different $\epsilon$ values on CIFAR10 and CelebA64.

\begin{table}[!ht]\small
\renewcommand{\arraystretch}{1.2}
\centering

\begin{tabular}{c|c|c|c|c}
\multirow{2}{*}{Privacy ($\epsilon)$} & \multicolumn{2}{c|}{CIFAR10} & 
\multicolumn{2}{c}{CelebA64}\\
\cline{2-5}
&{\sc PrivImage} & \system & {\sc PrivImage} & \system \\
\hline
1  &  29.8 & 63.2 &  71.4 & 60.5 \\
10  & 27.6 & 25.4 &  49.3 & 37.3 \\
\end{tabular}
\caption{Performance comparison of {\sc PrivImage} and \system (measured by FID scores). \label{tab:privimage}}
\end{table}

Notably, \system outperforms {\sc PrivImage} in most scenarios, with one exception: CIFAR10 under $\epsilon = 1$. This likely occurs because {\sc PrivImage} selects public data similar to the private data for pre-training. With clearly structured private data (for instance, CIFAR10 contains 10 distinct classes), using a targeted subset rather than all the public data tends to improve DP fine-tuning, especially under strict privacy budgets. However, this advantage may diminish with less structured private data (e.g., CelebA64). We consider leveraging the {\sc PrivImage}'s selective data approach to enhance \system as our ongoing research.

\subsection{Impact of Retrieval-Augmented Training}

\system can integrate with existing methods for training DP diffusion models since it is agnostic to model training, though its neighbor retrieval operates on latents, making it compatible only with latent diffusion models. To measure the impact of \system, we use a latent diffusion model as the backbone model for both \dpldm and \dpdm, evaluating their performance with and without RAPID. Table~\ref{tab:rgt} shows results on MNIST and CIFAR10 at $\epsilon = 10$. The substantial FID score improvement demonstrates the effectiveness of retrieval-augmented training.

\begin{table}[!ht]\small
\renewcommand{\arraystretch}{1.2}
\centering

\begin{tabular}{c|c|c|c|c}
\multirow{2}{*}{Dataset} & \multicolumn{2}{c|}{\dpdm} & 
\multicolumn{2}{c}{\dpldm}\\
\cline{2-5}
& w/o & w/ & w/o & w/ \\
\hline
MNIST  &42.9 & 25.4 & 27.2 & 14.1\\
CIFAR10  & 82.2 & 54.1 & 33.3 & 25.4\\
\end{tabular}
\caption{Impact of retrieval-augmented training on existing methods ($\epsilon = 10$). \label{tab:rgt}}
\end{table}

\subsection{Knowledge Base Generation}

While prior work on retrieval augmented generation (e.g., {\sc ReDi}~\cite{zhang2023redi}) requires all the trajectories to share the same latent, building the knowledge base needs to iteratively sample tens of thousands of trajectories from a pre-trained diffusion model (e.g., Stable Diffusion), which is highly expensive. For instance, on a workstation running one Nvidia RTX 6000 GPU, {\sc ReDi} requires over 8 hours to build a 10K-sample knowledge base.

\begin{table}[!ht]\small
\renewcommand{\arraystretch}{1.2}
\centering

\begin{tabular}{c|c|c|c|c|c|c|c}
Knowledge Base Size & 10K      &    20K      &  30K  &        40K &       50K    &     60K    &        70K\\
\hline
\system &  2.10s  &    4.17s  &   6.12s    &  8.23s   &  10.33s    &12.19s   &  14.48s\\
\end{tabular}
\caption{Runtime of \system for knowledge base construction. \label{tab:runtime}}
\end{table}

In comparison, \system eliminates this constraint, which allows it to directly compute the trajectory for each sample in the public dataset in a forward pass. Table~\ref{tab:runtime} shows \system's runtime efficiency for various knowledge base sizes, achieving orders of magnitude faster performance than prior work.

\subsection{Performance with Varying $\epsilon$}

\begin{figure}[!ht]
    \centering
    \includegraphics[width=1.0\linewidth]{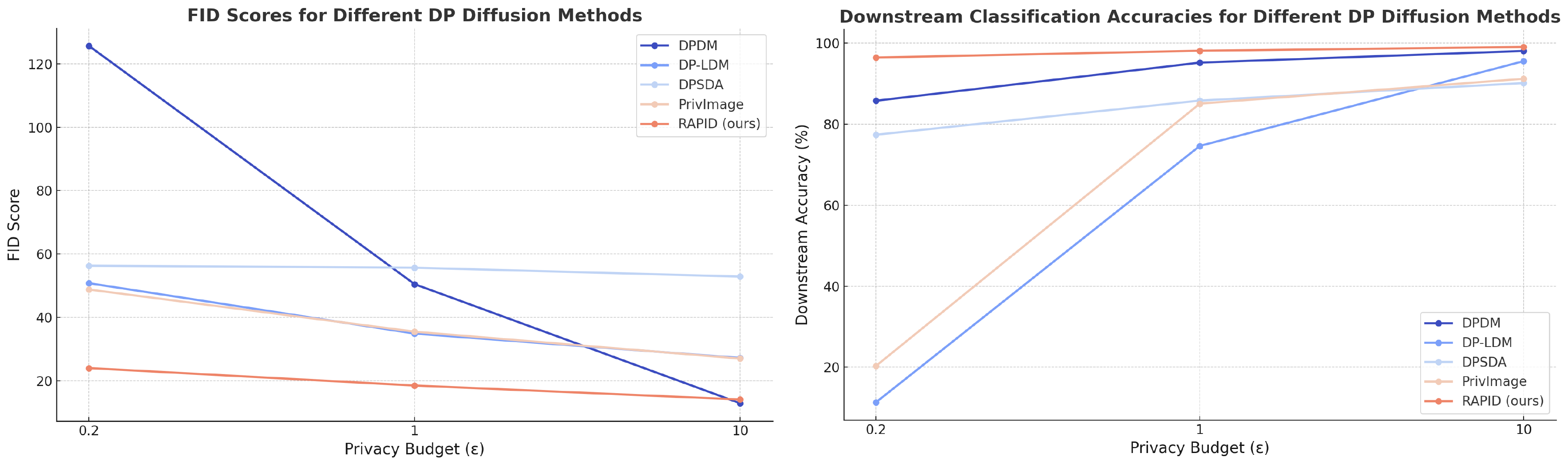}
      \caption{Performance of \system and baselines with varying $\epsilon$ on EMNIST$\rightarrow$MNIST: (a) FID scores and (b) downstream classification accuracy}
    \label{fig:epsilon-curve}
\end{figure}

Figure~\ref{fig:epsilon-curve} compares the performance (measured by FID scores and downstream classification accuracy) of \system and baselines (\dpdm,  \dpldm, \dpsda, and {\sc PrivImage}) in the case of EMNIST$\rightarrow$MNIST under varying $\epsilon$ settings. Observe that, under the same
privacy budget, \system considerably outperforms the baselines across most cases.

\section{Discussion}

\subsection{Comparison of \system and {\sc ReDi}}

{\sc ReDi}~\cite{zhang2023redi} also employs some strategies similar to \system such as constructing trajectory knowledge bases at early stages to bypass intermediate steps in the generation process. However, the two methods differ in several fundamental aspects. 

First, {\sc ReDi} employs RAG in the inference stage to improve generative efficiency, while \system integrates RAD into the DP training of diffusion models. Second, unlike {\sc ReDi} that builds its knowledge base by iteratively sampling tens of thousands of diffusion trajectories from a pre-trained latent diffusion model (e.g., Stable Diffusion), \system constructs the knowledge base by directly computing the diffusion trajectories via adding a scaled version of the initial latent to each sample in the public dataset, which greatly reduces the computational cost. Last, all the trajectories in {\sc ReDi} share the same initial latent. In contrast, the initial latents in  \system are randomly sampled, significantly improving the diversity of generated samples.



\subsection{Impact of Public/Private Data Similarity}

Like other DP diffusion model approaches (e.g., \dpdm, \dpldm, {\sc PrivImage}) and the broader pre-training/fine-tuning paradigm, \system assumes access to a diverse public dataset that captures a range of patterns. However, \system is more flexible: the public and private datasets need not closely match in distribution, as long as the public dataset contains similar high-level layouts. Here, we explore the possible explanations.

\begin{figure}[!ht]
    \centering
    \includegraphics[width=0.85\textwidth]{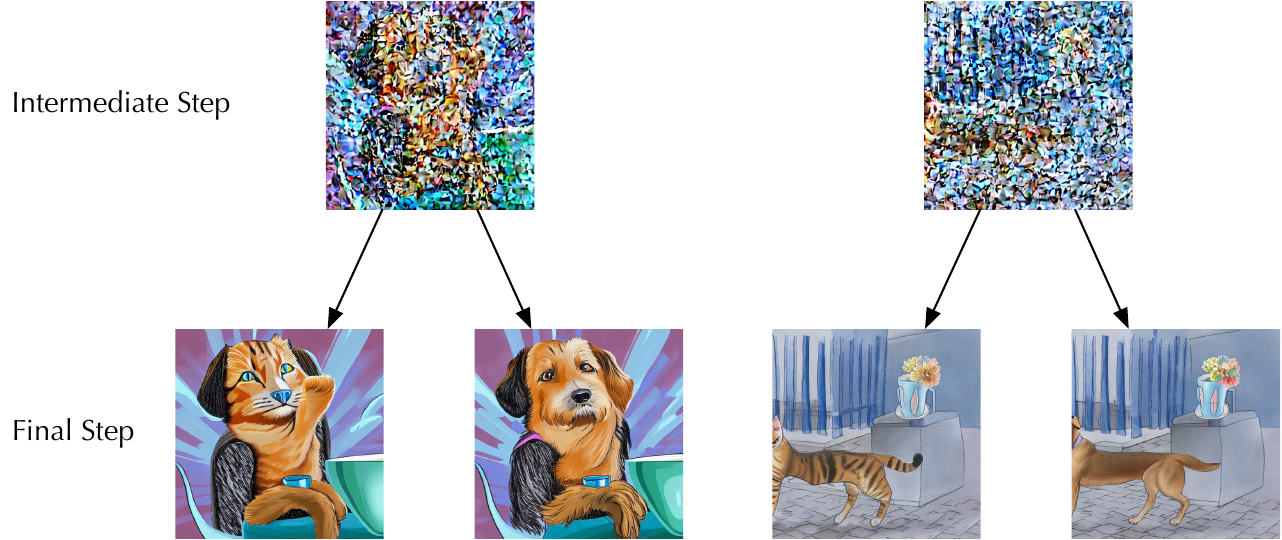}
    \caption{Disentanglement effects of diffusion models.}
    \label{fig:disentangle}
\end{figure}

Existing studies~\citep{sdedit,zhang2023redi} establish that in diffusion models, early stages determine image layouts that can be shared across many generation trajectories, while later steps define specific details. \cite{disentangle} further discover diffusion models' disentanglement capability, allowing generation of images with different styles and attributes from the same intermediate sampling stage, as shown in Figure~\ref{fig:disentangle}. This disentanglement property enables \system to maintain robust performance even when public and private dataset distributions differ significantly, provided their high-level layouts remain similar.

%% file: main.bbl
\begin{thebibliography}{57}
\providecommand{\natexlab}[1]{#1}
\providecommand{\url}[1]{\texttt{#1}}
\expandafter\ifx\csname urlstyle\endcsname\relax
  \providecommand{\doi}[1]{doi: #1}\else
  \providecommand{\doi}{doi: \begingroup \urlstyle{rm}\Url}\fi

\bibitem[Abadi et~al.(2016)Abadi, Chu, Goodfellow, McMahan, Mironov, Talwar, and Zhang]{abadi2016deep}
Martin Abadi, Andy Chu, Ian Goodfellow, H~Brendan McMahan, Ilya Mironov, Kunal Talwar, and Li~Zhang.
\newblock Deep learning with differential privacy.
\newblock In \emph{Proceedings of the ACM Conference on Computer and Communications (CCS)}, 2016.

\bibitem[Bar-Tal et~al.(2024)Bar-Tal, Chefer, Tov, Herrmann, Paiss, Zada, Ephrat, Hur, Li, Michaeli, et~al.]{bar2024lumiere}
Omer Bar-Tal, Hila Chefer, Omer Tov, Charles Herrmann, Roni Paiss, Shiran Zada, Ariel Ephrat, Junhwa Hur, Yuanzhen Li, Tomer Michaeli, et~al.
\newblock Lumiere: A space-time diffusion model for video generation.
\newblock \emph{ArXiv e-prints}, 2024.

\bibitem[Blattmann et~al.(2022)Blattmann, Rombach, Oktay, M{\"u}ller, and Ommer]{blattmann2022retrieval}
Andreas Blattmann, Robin Rombach, Kaan Oktay, Jonas M{\"u}ller, and Bj{\"o}rn Ommer.
\newblock Retrieval-augmented diffusion models.
\newblock In \emph{Proceedings of the Advances in Neural Information Processing Systems (NeurIPS)}, 2022.

\bibitem[Carlini et~al.(2023)Carlini, Hayes, Nasr, Jagielski, Sehwag, Tram\`{e}r, Balle, Ippolito, and Wallace]{carlini2023extracting}
Nicholas Carlini, Jamie Hayes, Milad Nasr, Matthew Jagielski, Vikash Sehwag, Florian Tram\`{e}r, Borja Balle, Daphne Ippolito, and Eric Wallace.
\newblock Extracting training data from diffusion models.
\newblock In \emph{Proceedings of the USENIX Security Symposium (SEC)}, 2023.

\bibitem[Casanova et~al.(2021)Casanova, Careil, Verbeek, Drozdzal, and Romero~Soriano]{casanova2021instance}
Arantxa Casanova, Marlene Careil, Jakob Verbeek, Michal Drozdzal, and Adriana Romero~Soriano.
\newblock Instance-conditioned gan.
\newblock In \emph{Proceedings of the Advances in Neural Information Processing Systems (NeurIPS)}, 2021.

\bibitem[Chen et~al.(2020{\natexlab{a}})Chen, Orekondy, and Fritz]{gs-wgan}
Dingfan Chen, Tribhuvanesh Orekondy, and Mario Fritz.
\newblock Gs-wgan: A gradient-sanitized approach for learning differentially private generators.
\newblock In \emph{Proceedings of the Advances in Neural Information Processing Systems (NeurIPS)}, 2020{\natexlab{a}}.

\bibitem[Chen et~al.(2020{\natexlab{b}})Chen, Kornblith, Norouzi, and Hinton]{chen2020simple}
Ting Chen, Simon Kornblith, Mohammad Norouzi, and Geoffrey Hinton.
\newblock A simple framework for contrastive learning of visual representations.
\newblock In \emph{Proceedings of the IEEE Conference on Machine Learning (ICML)}, 2020{\natexlab{b}}.

\bibitem[Chen et~al.(2020{\natexlab{c}})Chen, Fan, Girshick, and He]{chen2020improved}
Xinlei Chen, Haoqi Fan, Ross Girshick, and Kaiming He.
\newblock Improved baselines with momentum contrastive learning.
\newblock \emph{ArXiv e-prints}, 2020{\natexlab{c}}.

\bibitem[Cohen et~al.(2017)Cohen, Afshar, Tapson, and Van~Schaik]{cohen2017emnist}
Gregory Cohen, Saeed Afshar, Jonathan Tapson, and Andre Van~Schaik.
\newblock Emnist: Extending mnist to handwritten letters.
\newblock In \emph{Proceedings of International Joint Conference on Neural Networks (IJCNN)}, 2017.

\bibitem[Darlow et~al.(2018)Darlow, Crowley, Antoniou, and Storkey]{darlow2018cinic}
Luke~N Darlow, Elliot~J Crowley, Antreas Antoniou, and Amos~J Storkey.
\newblock Cinic-10 is not imagenet or cifar-10.
\newblock \emph{ArXiv e-prints}, 2018.

\bibitem[Deng et~al.(2009)Deng, Dong, Socher, Li, Li, and Fei-Fei]{deng2009imagenet}
Jia Deng, Wei Dong, Richard Socher, Li-Jia Li, Kai Li, and Li~Fei-Fei.
\newblock Imagenet: A large-scale hierarchical image database.
\newblock In \emph{Proceedings of the IEEE Conference on Computer Vision and Pattern Recognition (CVPR)}, 2009.

\bibitem[Deng(2012)]{deng2012mnist}
Li~Deng.
\newblock The mnist database of handwritten digit images for machine learning research.
\newblock \emph{IEEE Signal Processing Magazine}, 29\penalty0 (6):\penalty0 141--142, 2012.

\bibitem[Dockhorn et~al.(2023)Dockhorn, Cao, Vahdat, and Kreis]{dockhorn2022differentially}
Tim Dockhorn, Tianshi Cao, Arash Vahdat, and Karsten Kreis.
\newblock {Differentially Private Diffusion Models}.
\newblock \emph{Transactions on Machine Learning Research}, 2023.

\bibitem[Everingham(2005)]{voc}
Mark Everingham.
\newblock {The PASCAL Visual Object Classes Challenge 2005}, 2005.
\newblock URL \url{http://host.robots.ox.ac.uk/pascal/VOC/voc2005}.

\bibitem[Fuentes et~al.(2024)Fuentes, Mullins, McKenna, Miklau, and Sheldon]{fuentes2024joint}
Miguel Fuentes, Brett~C Mullins, Ryan McKenna, Gerome Miklau, and Daniel Sheldon.
\newblock Joint selection: Adaptively incorporating public information for private synthetic data.
\newblock In \emph{Proceedings of the International Conference on Artificial Intelligence and Statistics (AISTATS)}, 2024.

\bibitem[Ghalebikesabi et~al.(2023)Ghalebikesabi, Berrada, Gowal, Ktena, Stanforth, Hayes, De, Smith, Wiles, and Balle]{ghalebikesabi2023differentially}
Sahra Ghalebikesabi, Leonard Berrada, Sven Gowal, Ira Ktena, Robert Stanforth, Jamie Hayes, Soham De, Samuel~L Smith, Olivia Wiles, and Borja Balle.
\newblock Differentially private diffusion models generate useful synthetic images.
\newblock \emph{ArXiv e-prints}, 2023.

\bibitem[Goodfellow et~al.(2014)Goodfellow, Pouget-Abadie, Mirza, Xu, Warde-Farley, Ozair, Courville, and Bengio]{gan}
Ian~J. Goodfellow, Jean Pouget-Abadie, Mehdi Mirza, Bing Xu, David Warde-Farley, Sherjil Ozair, Aaron Courville, and Yoshua Bengio.
\newblock Generative adversarial networks.
\newblock In \emph{Proceedings of the Advances in Neural Information Processing Systems (NeurIPS)}, 2014.

\bibitem[{Gopi} et~al.(2021){Gopi}, {Tat Lee}, and {Wutschitz}]{privacy-accounting}
Sivakanth {Gopi}, Yin {Tat Lee}, and Lukas {Wutschitz}.
\newblock {Numerical Composition of Differential Privacy}.
\newblock In \emph{Proceedings of the Advances in Neural Information Processing Systems (NeurIPS)}, 2021.

\bibitem[Guu et~al.(2020)Guu, Lee, Tung, Pasupat, and Chang]{guu2020retrieval}
Kelvin Guu, Kenton Lee, Zora Tung, Panupong Pasupat, and Mingwei Chang.
\newblock Retrieval augmented language model pre-training.
\newblock In \emph{Proceedings of the IEEE Conference on Machine Learning (ICML)}, 2020.

\bibitem[Harder et~al.(2021)Harder, Adamczewski, and Park]{harder2021dpmerf}
Frederik Harder, Kamil Adamczewski, and Mijung Park.
\newblock Dp-merf: Differentially private mean embeddings with random features for practical privacy-preserving data generation.
\newblock \emph{Transactions on Machine Learning Research}, 2021.

\bibitem[Harder et~al.(2023)Harder, Jalali, Sutherland, and Park]{harder2023pretrained}
Frederik Harder, Milad Jalali, Danica~J. Sutherland, and Mijung Park.
\newblock Pre-trained perceptual features improve differentially private image generation.
\newblock \emph{Transactions on Machine Learning Research}, 2023.

\bibitem[He et~al.(2016)He, Zhang, Ren, and Sun]{he2016deep}
Kaiming He, Xiangyu Zhang, Shaoqing Ren, and Jian Sun.
\newblock Deep residual learning for image recognition.
\newblock In \emph{Proceedings of the IEEE Conference on Computer Vision and Pattern Recognition (CVPR)}, 2016.

\bibitem[Ho et~al.(2020)Ho, Jain, and Abbeel]{ho2020denoising}
Jonathan Ho, Ajay Jain, and Pieter Abbeel.
\newblock Denoising diffusion probabilistic models.
\newblock In \emph{Proceedings of the Advances in Neural Information Processing Systems (NeurIPS)}, 2020.

\bibitem[Hu et~al.(2023)Hu, Wu, Li, Long, Garrido, Ge, Ding, Forsyth, Li, and Song]{sok}
Yuzheng Hu, Fan Wu, Qinbin Li, Yunhui Long, Gonzalo~Munilla Garrido, Chang Ge, Bolin Ding, David Forsyth, Bo~Li, and Dawn Song.
\newblock Sok: Privacy-preserving data synthesis.
\newblock In \emph{Proceedings of the IEEE Symposium on Security and Privacy (S\&P)}, 2023.

\bibitem[Jiang et~al.(2022)Jiang, Zhang, Karami, Chen, Shao, and Yu]{jiang2022dp2vae}
Dihong Jiang, Guojun Zhang, Mahdi Karami, Xi~Chen, Yunfeng Shao, and Yaoliang Yu.
\newblock Dp$^2$-vae: Differentially private pre-trained variational autoencoders.
\newblock \emph{ArXiv e-prints}, 2022.

\bibitem[Karras et~al.(2019)Karras, Laine, and Aila]{karras2019style}
Tero Karras, Samuli Laine, and Timo Aila.
\newblock A style-based generator architecture for generative adversarial networks.
\newblock In \emph{Proceedings of the IEEE Conference on Computer Vision and Pattern Recognition (CVPR)}, 2019.

\bibitem[Khalil(2008)]{sensitivity}
Hassan~K. Khalil.
\newblock \emph{Nonlinear Systems Third Edition}.
\newblock Prentice Hall, 2008.

\bibitem[Khandelwal et~al.(2019)Khandelwal, Levy, Jurafsky, Zettlemoyer, and Lewis]{khandelwal2019generalization}
Urvashi Khandelwal, Omer Levy, Dan Jurafsky, Luke Zettlemoyer, and Mike Lewis.
\newblock Generalization through memorization: Nearest neighbor language models.
\newblock \emph{ArXiv e-prints}, 2019.

\bibitem[{Kingma} \& {Ba}(2014){Kingma} and {Ba}]{adam}
Diederik~P. {Kingma} and Jimmy {Ba}.
\newblock {Adam: A Method for Stochastic Optimization}.
\newblock \emph{ArXiv e-prints}, 2014.

\bibitem[Kong et~al.(2020)Kong, Ping, Huang, Zhao, and Catanzaro]{kong2020diffwave}
Zhifeng Kong, Wei Ping, Jiaji Huang, Kexin Zhao, and Bryan Catanzaro.
\newblock Diffwave: A versatile diffusion model for audio synthesis.
\newblock \emph{ArXiv e-prints}, 2020.

\bibitem[Krizhevsky et~al.(2009)Krizhevsky, Hinton, et~al.]{krizhevsky2009learning}
Alex Krizhevsky, Geoffrey Hinton, et~al.
\newblock Learning multiple layers of features from tiny images.
\newblock \emph{Master's thesis, Department of Computer Science, University of Toronto}, 2009.

\bibitem[Krizhevsky et~al.(2012)Krizhevsky, Sutskever, and Hinton]{krizhevsky2012imagenet}
Alex Krizhevsky, Ilya Sutskever, and Geoffrey~E Hinton.
\newblock Imagenet classification with deep convolutional neural networks.
\newblock In \emph{Proceedings of the Advances in Neural Information Processing Systems (NeurIPS)}, 2012.

\bibitem[Lebensold et~al.(2024)Lebensold, Sanjabi, Astolfi, Romero-Soriano, Chaudhuri, Rabbat, and Guo]{lebensold2024dp}
Jonathan Lebensold, Maziar Sanjabi, Pietro Astolfi, Adriana Romero-Soriano, Kamalika Chaudhuri, Mike Rabbat, and Chuan Guo.
\newblock Dp-rdm: Adapting diffusion models to private domains without fine-tuning.
\newblock \emph{ArXiv e-prints}, 2024.

\bibitem[Lewis et~al.(2020)Lewis, Perez, Piktus, Petroni, Karpukhin, Goyal, K{\"u}ttler, Lewis, Yih, Rockt{\"a}schel, et~al.]{lewis2020retrieval}
Patrick Lewis, Ethan Perez, Aleksandra Piktus, Fabio Petroni, Vladimir Karpukhin, Naman Goyal, Heinrich K{\"u}ttler, Mike Lewis, Wen-tau Yih, Tim Rockt{\"a}schel, et~al.
\newblock Retrieval-augmented generation for knowledge-intensive nlp tasks.
\newblock In \emph{Proceedings of the Advances in Neural Information Processing Systems (NeurIPS)}, 2020.

\bibitem[{Li} et~al.(2024){Li}, {Gong}, {Li}, {Zhao}, {Hou}, and {Wang}]{privimage}
Kecen {Li}, Chen {Gong}, Zhixiang {Li}, Yuzhong {Zhao}, Xinwen {Hou}, and Tianhao {Wang}.
\newblock {PrivImage: Differentially Private Synthetic Image Generation using Diffusion Models with Semantic-Aware Pretraining}.
\newblock In \emph{Proceedings of the USENIX Security Symposium (SEC)}, 2024.

\bibitem[Liew et~al.(2022)Liew, Takahashi, and Ueno]{liew2022pearl}
Seng~Pei Liew, Tsubasa Takahashi, and Michihiko Ueno.
\newblock Pearl: Data synthesis via private embeddings and adversarial reconstruction learning.
\newblock In \emph{Proceedings of the International Conference on Learning Representations (ICLR)}, 2022.

\bibitem[{Lin} et~al.(2024){Lin}, {Gopi}, {Kulkarni}, {Nori}, and {Yekhanin}]{dpsda}
Zinan {Lin}, Sivakanth {Gopi}, Janardhan {Kulkarni}, Harsha {Nori}, and Sergey {Yekhanin}.
\newblock {Differentially Private Synthetic Data via Foundation Model APIs}.
\newblock In \emph{Proceedings of the International Conference on Learning Representations (ICLR)}, 2024.

\bibitem[Liu et~al.(2022)Liu, Ren, Lin, and Zhao]{liu2022pseudo}
Luping Liu, Yi~Ren, Zhijie Lin, and Zhou Zhao.
\newblock Pseudo numerical methods for diffusion models on manifolds.
\newblock In \emph{Proceedings of the International Conference on Learning Representations (ICLR)}, 2022.

\bibitem[Liu et~al.(2021{\natexlab{a}})Liu, Vietri, Steinke, Ullman, and Wu]{liu2021leveraging}
Terrance Liu, Giuseppe Vietri, Thomas Steinke, Jonathan Ullman, and Steven Wu.
\newblock Leveraging public data for practical private query release.
\newblock In \emph{Proceedings of the IEEE Conference on Machine Learning (ICML)}, 2021{\natexlab{a}}.

\bibitem[Liu et~al.(2021{\natexlab{b}})Liu, Vietri, and Wu]{liu2021iterative}
Terrance Liu, Giuseppe Vietri, and Steven~Z Wu.
\newblock Iterative methods for private synthetic data: Unifying framework and new methods.
\newblock In \emph{Proceedings of the Advances in Neural Information Processing Systems (NeurIPS)}, 2021{\natexlab{b}}.

\bibitem[Liu et~al.(2015)Liu, Luo, Wang, and Tang]{liu2015faceattributes}
Ziwei Liu, Ping Luo, Xiaogang Wang, and Xiaoou Tang.
\newblock Deep learning face attributes in the wild.
\newblock In \emph{Proceedings of the IEEE International Conference on Computer Vision (ICCV)}, 2015.

\bibitem[Lyu et~al.(2023)Lyu, Vinaroz, Liu, and Park]{lyu2023differentially}
Saiyue Lyu, Margarita Vinaroz, Michael~F Liu, and Mijung Park.
\newblock Differentially private latent diffusion models.
\newblock \emph{ArXiv e-prints}, 2023.

\bibitem[{Meng} et~al.(2021){Meng}, {He}, {Song}, {Song}, {Wu}, {Zhu}, and {Ermon}]{sdedit}
Chenlin {Meng}, Yutong {He}, Yang {Song}, Jiaming {Song}, Jiajun {Wu}, Jun-Yan {Zhu}, and Stefano {Ermon}.
\newblock {SDEdit: Guided Image Synthesis and Editing with Stochastic Differential Equations}.
\newblock In \emph{Proceedings of the International Conference on Learning Representations (ICLR)}, 2021.

\bibitem[{Mironov}(2017)]{renyi}
Ilya {Mironov}.
\newblock {Renyi Differential Privacy}.
\newblock In \emph{Proceedings of the IEEE Computer Security Foundations Symposium (CSF)}, 2017.

\bibitem[{Mironov} et~al.(2019){Mironov}, {Talwar}, and {Zhang}]{renyi-gm}
Ilya {Mironov}, Kunal {Talwar}, and Li~{Zhang}.
\newblock {R{\'e}nyi Differential Privacy of the Sampled Gaussian Mechanism}.
\newblock \emph{ArXiv e-prints}, 2019.

\bibitem[Naeem et~al.(2020)Naeem, Oh, Uh, Choi, and Yoo]{naeem2020reliable}
Muhammad~Ferjad Naeem, Seong~Joon Oh, Youngjung Uh, Yunjey Choi, and Jaejun Yoo.
\newblock Reliable fidelity and diversity metrics for generative models.
\newblock In \emph{Proceedings of the IEEE Conference on Machine Learning (ICML)}, 2020.

\bibitem[Oord et~al.(2018)Oord, Li, and Vinyals]{oord2018representation}
Aaron van~den Oord, Yazhe Li, and Oriol Vinyals.
\newblock Representation learning with contrastive predictive coding.
\newblock \emph{ArXiv e-prints}, 2018.

\bibitem[Rombach et~al.(2022)Rombach, Blattmann, Lorenz, Esser, and Ommer]{rombach2022high}
Robin Rombach, Andreas Blattmann, Dominik Lorenz, Patrick Esser, and Bj{\"o}rn Ommer.
\newblock High-resolution image synthesis with latent diffusion models.
\newblock In \emph{Proceedings of the IEEE Conference on Computer Vision and Pattern Recognition (CVPR)}, 2022.

\bibitem[Song et~al.(2020)Song, Meng, and Ermon]{song2020denoising}
Jiaming Song, Chenlin Meng, and Stefano Ermon.
\newblock Denoising diffusion implicit models.
\newblock \emph{ArXiv e-prints}, 2020.

\bibitem[Song et~al.(2021)Song, Sohl-Dickstein, Kingma, Kumar, Ermon, and Poole]{song2021scorebased}
Yang Song, Jascha Sohl-Dickstein, Diederik~P Kingma, Abhishek Kumar, Stefano Ermon, and Ben Poole.
\newblock Score-based generative modeling through stochastic differential equations.
\newblock In \emph{Proceedings of the International Conference on Learning Representations (ICLR)}, 2021.

\bibitem[Vinaroz et~al.(2022)Vinaroz, Charusaie, Harder, Adamczewski, and Park]{vinaroz2022hermite}
Margarita Vinaroz, Mohammad-Amin Charusaie, Frederik Harder, Kamil Adamczewski, and Mijung Park.
\newblock Hermite polynomial features for private data generation.
\newblock In \emph{Proceedings of the IEEE Conference on Machine Learning (ICML)}, 2022.

\bibitem[Wang et~al.(2024)Wang, Pang, Lu, Rao, Zhou, and Xue]{wangdp}
Haichen Wang, Shuchao Pang, Zhigang Lu, Yihang Rao, Yongbin Zhou, and Minhui Xue.
\newblock Dp-promise: Differentially private diffusion probabilistic models for image synthesis.
\newblock In \emph{Proceedings of the USENIX Security Symposium (SEC)}, 2024.

\bibitem[Wen et~al.(2023)Wen, Liu, Chen, and Lyu]{wen2023detecting}
Yuxin Wen, Yuchen Liu, Chen Chen, and Lingjuan Lyu.
\newblock Detecting, explaining, and mitigating memorization in diffusion models.
\newblock In \emph{Proceedings of the International Conference on Learning Representations (ICLR)}, 2023.

\bibitem[{Wu} et~al.(2022){Wu}, {Liu}, {Zhao}, {Kale}, {Bui}, {Yu}, {Lin}, {Zhang}, and {Chang}]{disentangle}
Qiucheng {Wu}, Yujian {Liu}, Handong {Zhao}, Ajinkya {Kale}, Trung {Bui}, Tong {Yu}, Zhe {Lin}, Yang {Zhang}, and Shiyu {Chang}.
\newblock {Uncovering the Disentanglement Capability in Text-to-Image Diffusion Models}.
\newblock In \emph{Proceedings of the IEEE Conference on Computer Vision and Pattern Recognition (CVPR)}, 2022.

\bibitem[Yoon et~al.(2019)Yoon, Jordon, and van~der Schaar]{pate-gan}
Jinsung Yoon, James Jordon, and Mihaela van~der Schaar.
\newblock {PATE}-{GAN}: Generating synthetic data with differential privacy guarantees.
\newblock In \emph{Proceedings of the International Conference on Learning Representations (ICLR)}, 2019.

\bibitem[{Yousefpour} et~al.(2021){Yousefpour}, {Shilov}, {Sablayrolles}, {Testuggine}, {Prasad}, {Malek}, {Nguyen}, {Ghosh}, {Bharadwaj}, {Zhao}, {Cormode}, and {Mironov}]{opacus}
Ashkan {Yousefpour}, Igor {Shilov}, Alexandre {Sablayrolles}, Davide {Testuggine}, Karthik {Prasad}, Mani {Malek}, John {Nguyen}, Sayan {Ghosh}, Akash {Bharadwaj}, Jessica {Zhao}, Graham {Cormode}, and Ilya {Mironov}.
\newblock {Opacus: User-Friendly DockhornDifferential Privacy Library in PyTorch}.
\newblock \emph{ArXiv e-prints}, 2021.

\bibitem[Zhang et~al.(2023)Zhang, Yang, Wang, and Li]{zhang2023redi}
Kexun Zhang, Xianjun Yang, William~Yang Wang, and Lei Li.
\newblock Redi: efficient learning-free diffusion inference via trajectory retrieval.
\newblock In \emph{Proceedings of the IEEE Conference on Machine Learning (ICML)}, 2023.

\end{thebibliography}
